\documentclass[12pt,a4paper]{article}
\usepackage[a4paper]{geometry}

\usepackage{amsmath}
\usepackage{amsfonts}
\usepackage{amssymb}
\usepackage{dsfont}

\usepackage{verbatim}
\usepackage{tikz}
\usepackage{cite} 
\usepackage{hyperref}

\allowdisplaybreaks[2]
\numberwithin{equation}{section}

\newcommand{\eq}[1]{\begin{equation}
                     \begin{split} #1 \end{split}
                     \end{equation}}

\newcommand{\op}{\hspace{1pt}}

\newcommand{\be}{\begin{equation}}
\newcommand{\ee}{\end{equation}}

\newcommand{\beq}{\begin{equation}}
\newcommand{\eeq}{\end{equation}}

\newcommand{\cO}{\mathcal{O}}

\newcommand{\cE}{\mathcal{E}}

\newcommand{\cK}{\mathcal{K}}
\newcommand{\cM}{\mathcal{M}}

\newcommand{\cV}{\mathcal{V}}

\newcommand{\bbZ}{\mathbb{Z}}

\newcommand{\lv}{\ell}
\newcommand{\hG}{\hat G}
\newcommand{\Vell}{V_{\boldsymbol{\ell}}}

\newcommand{\Vh}{V_{\mathrm{heavy}}}
\newcommand{\Vl}{V_{\mathrm{light}}}
\newcommand{\Vr}{V_{\mathrm{rest}}}

\begin{document}
\pagestyle{empty}


\begin{center}

\vspace*{1em}

{\LARGE\bf The Tadpole Conjecture\\ in Asymptotic Limits\\}

\vspace*{3em}

\large{Mariana Gra\~na$^{\flat}$, Thomas W. Grimm$^{\sharp}$, Damian van de Heisteeg$^{\sharp}$,
\\ [0.2mm] Alvaro Herraez$^{\flat}$  and Erik Plauschinn$^{\sharp}$\\}

\vspace*{1.5em}

{\small $^\flat$  Institut de Physique Th\'eorique, Universit\'e Paris Saclay, CEA, CNRS\\ }
{\small\it  Orme des Merisiers, 91191 Gif-sur-Yvette CEDEX, France.\\} 
\vspace*{0.5em}
{\small  $^\sharp$ Institute for Theoretical Physics, Utrecht University\\}
{\small\it 3584 CC Utrecht, The Netherlands\\}

\vspace*{1.5em}

{\small \verb"mariana.grana@ipht.fr, t.w.grimm@uu.nl, d.t.e.vandeheisteeg@uu.nl, "\\
\small \verb"alvaro.herraezescudero@ipht.fr, e.plauschinn@uu.nl"\\}

\vspace*{3em}

\small{\bf Abstract} \\[3mm]
\end{center}
The tadpole conjecture suggests that the complete stabilization of  complex structure deformations in Type IIB and F-theory flux 
compactifications is severely obstructed by the tadpole bound on the fluxes. More precisely, it states that the stabilization of  
a large number of moduli requires a flux background with a tadpole that scales linearly in the number of stabilized fields. Restricting 
to the asymptotic regions of the complex structure moduli space, we give the first conceptual argument that explains this 
linear scaling setting and clarifies why it sets in only for a large number of stabilized moduli. Our approach relies on the use of asymptotic Hodge theory. 
In particular, we use the fact that in each asymptotic regime an orthogonal 
sl(2)-block structure emerges that allows us to group fluxes into sl(2)-representations 
and decouple complex structure directions. We show that the number of stabilized moduli scales with 
the number of sl(2)-representations supported by fluxes, and that each representation fixes a single modulus. Furthermore, we find that for 
Calabi-Yau four-folds all but one representation can be identified 
with representations occurring on two-folds. This allows us to discuss moduli stabilization 
explicitly and establish the relevant scaling constraints for the tadpole.


\clearpage

\setcounter{page}{1}
\pagestyle{plain}
\tableofcontents


\clearpage

\section{Introduction}

String theory compactifications typically give rise to a large number of massless scalar fields  in 
the lower-dimensional effective theory. These (moduli) fields correspond to 
deformations of the compactification background and  are ruled-out by experiments.
A very well-studied setting is that of Calabi-Yau orientifold compactifications of type II string theories, in particular, type IIB or its non-perturbative completion F-theory. The reason why this corner of the landscape is special is because one can turn on 
three-form fluxes (or four-form fluxes in the F/M-theory realisation) such that their back-reaction still allows for a Calabi-Yau geometry, as long as they are self-dual \cite{Becker:1996gj,Dasgupta:1999ss,Grana:2000jj}.  Furthermore, these fluxes generically lift the moduli corresponding to complex structure deformations of the Calabi-Yau manifold \cite{Dasgupta:1999ss,Giddings:2001yu}, and have therefore been used at length in string phenomenology.

The in general complicated structure
of the moduli space of complex structure deformations suggests that turning on flux quanta in a few cycles is enough to generate a potential capable of 
stabilizing a large number of moduli. However, this reasoning has been challenged by the ``Tadpole Conjecture" \cite{Bena:2020xrh,Bena:2021wyr}, 
which states that the charge $Q$ induced by the fluxes  needed to stabilize a large number, $n$, of moduli grows linearly with $n$. This growth yields 
an obstruction on complete moduli stabilization if its slope is larger than the slope for the linear growth observed for the tadpole. 
The refined version of the tadpole conjecture makes precisely this claim.
If this refined version of the tadpole conjecture is true, then the only phenomenologically-relevant Calabi-Yau manifolds are those with a small number of complex structure moduli, since large numbers are not amenable to flux-moduli stabilisation. While the tadpole conjecture suggests a mathematically precise statement, its proof is extremely 
challenging and currently wide open. Even in specific situations, such as~e.g.~K3$\times$K3 compactifications, any direct proof  soon 
faces technical challenges. 

In this work we initiate the 
first systematic and general approach to establish the tadpole conjecture.
We collect strong evidence that this conjecture is satisfied for all (strict) asymptotic limits in moduli space. Moreover, we give the first 
conceptual argument that explains the linear scaling of the tadpole with the number of stabilized moduli and the requirement that one has to consider a large number of moduli. 
Our approach uses the powerful machinery of asymptotic Hodge 
structures \cite{Schmid,CKS} and does neither rely on explicit examples nor on a specific choice of asymptotic limit. Rather, we exploit the fact that  
asymptotic Hodge theory provides us with an explicit expression for the Hodge-star operator acting on 
four-forms in terms of $n$ moduli that are taken to be in the asymptotic region.  Requiring the self-duality condition for the fluxes then becomes an 
explicit, polynomial equation for the moduli that has been explored before in \cite{Grimm:2019ixq,Grimm:2021ckh}. 
The degrees of these polynomials are fixed by the weights of the flux components under $n$ commuting sl(2)-algebras characterizing the asymptotic region.  
We show that there is a one-to-one correspondence between the number of moduli fixed and the number of sl(2)-representations supported by flux.  
Each representation yields an independent positive-definite term in the tadpole cancellation condition such that a scaling of the flux tadpole with 
the number of moduli appears to be immediate if the individual terms are lower-bounded. 

The crucial task of this work is to establish this scaling of the flux charge and argue for a lower bound on individual 
terms arising from fluxes in an sl(2)-representation, thus showing that the refined Tadpole Conjecture applies. This requires us to understand in detail which sl(2)-representations can arise in an asymptotic Calabi-Yau four-fold compactification. We find that the richest structure 
hereby comes from the presence of the single (4,0)-form, which accounts for sl(2)-representations with weights reaching from 
$-4$ to $4$. The remaining other sl$(2)$-representations have a maximal weight $2$ and match with sl$(2)$-representations 
found for Calabi-Yau two-folds (K3 surfaces). Since only the number of these K3-type representations
is increasing with the number of moduli, we are lead to study the much simpler and more constrained problem of examining 
the tadpole contribution of such shorter representations. We find that fluxes in these representations have 
either all $n$ sl$(2)$-weights greater or equal to zero, or smaller or equal to zero. The self-duality condition  
furthermore forbids purely zero-weight fluxes and relates fluxes with positive and negative weights. Any asymptotic 
stabilization thus has to use these K3-type flux pairs which individually fix a single modulus and contribute to the tadpole with 
positive powers of fractions of moduli that become large in the asymptotic regime. Remarkably, this tadpole can be lower-bounded 
by a moduli-independent sum of positive terms depending on the number sl(2)-fluxes times a large parameter ensuring that the moduli are evaluated in the asymptotic region.\footnote{This parameter, called $\gamma$, does not need to be extremely large. It was shown that already for $\gamma \gtrsim 4 $ moduli stabilization using asymptotic Hodge theory gives a very good approximation of the situation, where all corrections are taken into account \cite{Grimm:2021ckh}.} We complete our argument by 
providing evidence that the flux terms do not  scale inversely with the number of moduli and most of them are bounded from below by 1/4 in all examples we worked out.

The paper is organized as follows: in Section~\ref{sec_f_theory_flux} we introduce the basic features of F-theory flux compactifications and  the Tadpole Conjecture. In Section~\ref{sec:AHT} we discuss asymptotic Hodge theory and show that there are only very few weights that appear in a generic four-fold.  In Section~\ref{sec_modstab_gen} we show how moduli stabilization by self-dual fluxes works in general, while in Section~\ref{sec_modstab_explicitly} we work out the explicit solution for the representations appearing in a four-fold.  In Section~\ref{sec:tadpolecontribution} we analyze the tadpole, and show how the tadpole conjecture is verified. We conclude in Section~\ref{sec:conclusions}. In Appendix  \ref{app:AHS} we provide more details about asymptotic Hodge structures, and in Appendix \ref{app:Cinfty} we give explicit bases for the sl(2)-representations and the action of the Hodge star on them. In Appendix \ref{app:example} we illustrate the concepts introduced with a two-moduli example, while in Appendix \ref{app_flux_quant} we work out examples that illustrate flux quantization in the sl(2) basis with a large number of moduli.


\section{F-theory fluxes and the tadpole conjecture}
\label{sec_f_theory_flux}

In this section we give a brief review of flux compactifications in F-theory. 
We focus on the aspects needed for our study of the tadpole conjecture \cite{Bena:2020xrh,Bena:2021wyr},
and refer for a more comprehensive discussion to \cite{Denef:2008wq}. 


\subsection{F-theory flux compactifications}

We are interested in compactifications of F-theory on Calabi-Yau four-folds in the presence 
of four-form fluxes. Such manifolds  have to admit a two-torus fibration, which from the Type IIB 
perspective encodes  the variation of the dilaton-axion field. To study the resulting four-dimensional effective action we first consider the 
dual M-theory setting by compactifying eleven-dimensional supergravity on a resolved four-fold $Y_4$
to obtain a three-dimensional effective theory. To connect with F-theory one then takes the limit 
of shrinking the torus-fiber \cite{Denef:2008wq,Grimm:2010ks}.\footnote{Note that we consider the general 
case in which the F-theory setting can admit non-trivial 7-brane configurations with non-Abelian gauge groups. In this 
case $Y_4$ is the smooth Calabi-Yau four-fold obtained by resolving the gauge theory singularities. The volumes 
of the resolution cycles are parameterized by K\"ahler moduli and are shrunk in the F-theory limit.} Investigating the impact of corrections in the resulting F-theory effective actions has been the focus of \cite{Grimm:2013gma,Grimm:2014efa,Grimm:2015mua,Minasian:2015bxa,Weissenbacher:2019mef, Klaewer:2020lfg, Cicoli:2021rub, Gao:2022fdi}.

In the three-dimensional theory
and in the absence of fluxes the deformations of the four-fold give rise to
massless scalar fields,
which correspond to the 
$h^{3,1}$ complex-structure moduli and $h^{1,1}$ K\"ahler moduli of $Y_4$.\footnote{All the $h^{3,1}$ complex structure moduli lift directly to complex scalars in the four-dimensional F-theory lift, whereas only $h^{1,1}-1$ K\"ahler moduli, corresponding to the $(1,1)$-cohomology on the base, give rise to K\"ahler moduli in four dimensions.}
However, when considering a non-trivial four-form flux  
$G_4 \in H^4(Y_4,\bbZ/2)$,
a potential is induced which can be brought into the form \cite{Haack:2001jz,Denef:2008wq}
\begin{equation}\label{eq:potential}
V = \frac{1}{\cV^3}\bigl( \lVert G_4 \rVert^2 - \langle G_4 , G_4 \rangle \bigr)\,.
\end{equation}
Here $\cV $ denotes the volume of the four-fold $Y_4$ which depends on the K\"ahler moduli, 
and we have defined the norm and inner product 
of a real four-form $v\in H^4(Y_4,\mathbb R)$ as
\beq \label{norm&ip}
\lVert v\rVert^2=\int_{Y_4} v \wedge \star v \ , \hspace{50pt} \langle v,v\rangle =\int_{Y_4} v \wedge  v  \,,
\eeq
where $\star$ denotes the Hodge-star operator of $Y_4$. 
The dependence of the potential \eqref{eq:potential} on the complex-structure moduli is encoded 
in the Hodge-star operator in the first term, while
the second term is an on-shell contribution obtained using the Bianchi identity for $G_4$.

We are interested in Minkowski minima of the potential \eqref{eq:potential}. 
These are obtained when  $G_4$ is self-dual and primitive \cite{Dasgupta:1999ss,Becker:1996gj}, which
reads in formulas 
\eq{
  \label{cond_001}
  G_4 = \star G_4 \,, \hspace{50pt} J\wedge G_4 = 0\,.
}
Let us emphasize that in the following we consider $G_4 = \star G_4$ as a condition in cohomology and not as a local condition in target space.
We furthermore note that the primitive four-form cohomology decomposes as
\beq
\label{h4_decomp}
H^4_{\rm prim}=H^{4,0} \oplus H^{3,1} \oplus H_{\rm prim}^{2,2} \oplus H^{1,3} \oplus H^{0,4} \,,
\eeq
and that the Hodge-star operator $\star$ acts on the $H^{p,q}$ cohomologies by multiplication with $(-1)^{(p-q)/2}$. 
The self-duality condition in \eqref{cond_001} then  implies that the $(3,1)$-components of the 
four-form flux have to vanish, which  can be written as 
$h^{3,1}$ equations for the $h^{3,1}$ complex structure moduli. 
Generically these equations fix the moduli at some value -- but the tadpole conjecture, 
to which we turn now, challenges this na\"ive expectation.


\subsection{Tadpole conjecture}
\label{ss:tadpoleconjecture}

The four-form flux $G_4$ induces a D3-brane charge $Q$ (or M2-brane charge in the M-theory dual picture) 
that has to be cancelled globally. Defining 
\eq{
\label{charge_q}
Q=\frac12 \op\langle G_4 , G_4 \rangle  = \frac12 \int_{Y_4} G_4 \wedge G_4\,,
}
this leads to the tadpole cancelation condition and inequality
\beq
\label{Ftadpole}
Q+ N_{\rm D3}= \frac{\chi(Y_4)}{24} 
\hspace{30pt}\Rightarrow\hspace{30pt}
Q \leq \frac{\chi(Y_4)}{24} \,,
\eeq
where $\chi(Y_4)$ is the Euler number of $Y_4$ and $N_{\rm D3}\geq0$ denotes the 
number of space-time filling D3-(or M2-)branes. 
For solutions to the equations of motion with self-dual $G_4$ flux the charge $Q$ is always positive.
The fact that $Q$ is also bounded from above may suggest that the number of self-dual flux configurations 
on a given Calabi-Yau manifold is finite. However, as we will discuss in detail below, the Hodge star can degenerate 
near the boundaries of the moduli space and vacua could accumulate in such asymptotic regimes \cite{Ashok:2003gk}.
Remarkably, it turns out that one can prove the absence of such accumulation points and a general finiteness theorem
for self-dual flux vacua using asymptotic Hodge theory and tame geometry \cite{Bakker:2021uqw}.\footnote{This result 
is a generalization of a famous theorem \cite{CattaniDeligneKaplan} about the finiteness of Hodge loci, and captures finiteness along 
one-dimensional limits first shown in \cite{Schnell,Grimm:2020cda}.} 
Although  the number of flux configurations is
finite, it is expected to be extremely large. For instance, for the Calabi-Yau four-fold with the largest-known Euler number the number of 
vacua has been estimated to be of order $10^{272000}$ \cite{Taylor:2015xtz}. We note, though, that this number was 
obtained by counting lattice sites and does not take into account any of the intricate structure of the complex-structure moduli 
space and of the Hodge-star operator. The tadpole conjecture scrutinizes these estimates even further as it challenges the idea 
that full moduli stabilization can be achieved for manifolds with large $h^{3,1}$.

The tadpole conjecture \cite{Bena:2020xrh} postulates that for a large number of moduli there are always remaining flat directions. To be more precise, the conjecture states  that 
when stabilizing a large number of complex moduli $n_{\rm stab}\gg 1$, 
the charge induced by the flux grows linearly with the moduli in the form
\beq
Q > \alpha \,  n_{\rm stab}\,. 
\eeq
The refined version of the conjecture then gives a precise lower bound for the slope:
\beq \label{refined}
\alpha > \frac{1}{3} \, .
\eeq
If the tadpole conjecture is true, one cannot stabilize a large number of complex structure moduli within the tadpole bound 
\eqref{Ftadpole}, since for a large $h^{3,1}$ the Euler number behaves as 
$\chi(Y_4) \sim h^{3,1}/4$ and one would have
\eq{
 \frac13 n_{\rm stab} < Q\le \frac{\chi(Y_4)}{24} \sim \frac14\op h^{3,1} 
 \hspace{30pt}  \Rightarrow \hspace{30pt} 
 n_{\rm stab} < \frac34 h^{3,1} \, .
}
We note that the tadpole conjecture refers to stabilization of the real and imaginary part of the complex moduli. If the  conjecture is true there is always some left-over moduli space, and it is a very interesting question (beyond the scope of this paper) to understand its structure, in particular, if it is compact or not.


\section{Aspects of asymptotic Hodge theory}
\label{sec:AHT}

In this work we are interested in the behavior of the flux-induced charge $Q$ 
when stabilizing moduli near the 
boundary of  complex-structure moduli space. 
A suitable framework for discussing this question is asymptotic Hodge theory, which 
we briefly review in the following \cite{Grimm:2019ixq}.
Let us stress that a
boundary in complex-structure moduli space corresponds not only to the familiar 
large complex-struc\-ture limit, but it includes also the conifold point and more general degenerations \cite{Bastian:2021eom}.
Our analysis is valid for all of such boundaries. In Appendix \ref{app:example} we present a two-moduli example that illustrates the concepts we introduce in this section, where the asymptotic region we explore is that close to the conifold point and the weak coupling limit.


\subsection{Strict asymptotic regimes and sl(2)-decomposition}

The complex-structure moduli space of Calabi-Yau four-folds is parametrized by 
$h^{3,1}$ complex scalar fields. Let us consider  an asymptotic region in this moduli 
space and separate these fields into two groups in the following way
\eq{
  \{t^i,\zeta^{\alpha} \}\,,
  \hspace{50pt}
  \arraycolsep2pt
  i = 1,\ldots, n\,,
  \qquad
  \alpha = n+1, \ldots, h^{3,1} \,,
}
where $0<n\leq h^{3,1}$.
The $\zeta^{\alpha}$ are called \textit{spectator} fields and will not play any role 
in our subsequent discussion, so we will mostly ignore them.
The scalars $t^i$ correspond to the coordinates (on the covering space) of the moduli space that parametrize 
how far away we are from one of its boundaries. The real and imaginary parts of $t^i$ will loosely be called axions and saxions, respectively,\footnote{Note that the real parts are not necessarily axions, i.e.~they might not have a continuous shift symmetries even in the leading moduli space metric, such as the intersection of two conifold loci \cite{Bastian:2021eom}.} 
\eq{
  t^i = \phi^i + i\, s^i \,, 
  \hspace{30pt} s^i>0\,,
  \hspace{50pt} i=1,\ldots, n \, ,
}
and the boundary of the moduli space corresponds to the limit $s^i\to\infty$.


\subsubsection*{Asymptotic regimes}

In the near-boundary region we can consider the following two regimes:
\begin{enumerate}

\item The \textit{asymptotic regime} is characterized by the following condition for  the saxions $s^i$
with $i = 1,\ldots,n$
\eq{
s^i \gg 1 \, ,
} 
which  corresponds to dropping corrections ${\cal O}(e^{2 \pi i t^i})$ in, 
for instance, the Hodge-star operator.

\item In the \textit{strict asymptotic regime} the $s^i$ are ordered according to the hierarchy
\eq{
  \label{sector_001}
 \frac{s^1}{s^2}>\gamma \,, \quad  \frac{s^2}{s^3}>\gamma \, , \quad  \ldots \, , \quad 
  \frac{s^{n-1}}{s^{n}}>\gamma \, , \quad s^{n}>\gamma \, , \hspace{30pt}
  \lvert\phi^{ i}\rvert<\delta\,,
}
where $\gamma\gg1$ and $\delta>0$. Dropping polynomial corrections of order $\cO(\gamma^{-1})$ corresponds to the sl(2)-approximation \cite{CKS}, which is the setting where our work will take place. Note that these inequalities specify a certain hierarchy of field values and one 
cannot simply permute indices. 
In general, using the same coordinates with a different ordering, and hence a different 
hierarchy, implies that one probes another strict asymptotic regime with different properties.

\end{enumerate}
Let us emphasize that the spectator fields  $\zeta^\alpha$ are not send to the 
boundary, however, 
quantities such as the Hodge-star operator can still depend on them.


\subsubsection*{Sl(2)-decomposition}

We now turn to the sl(2)-decomposition of the fourth cohomology of 
$Y_4$. A pedagogical introduction to this subject can be found 
for instance in section~3.1 of \cite{Grimm:2021ckh}, and more detailed
discussions can be found in \cite{
Schmid, 
CKS,
Grimm:2018ohb,
Grimm:2018cpv}. 
We summarize the main aspects that we will use in the rest of the paper as follows:
\begin{itemize}

\item To each boundary of complex-structure moduli space one can associate monodromy
transformations acting on the fourth cohomology of the Calabi-Yau four-fold $Y_4$. 
If $n$  saxions $s^{i}$ are sent to the boundary, this action can be realized 
by $n$ commuting matrices $T_{i}$ acting on $H^4_{\rm prim}(Y_4, \mathbb R)$. 
These matrices can always be made
unipotent, that is $(T_{ i} - \mathds 1)^{m+1}=0$ for some $m\geq 0$,
and to each $T_{ i}$ we can associate a so-called log-monodromy matrix
$N_{ i} = \log T_{i}$.
Since the $T_{i}$ are unipotent it follows that the $N_{i}$ are nilpotent
(when acting on the fourth cohomology), 
and we note that the $N_{i}$ commute among each other.

\item Each nilpotent $N_i$ can be completed into an sl(2)-triple, in which it acts as lowering operator. However, the choice of weight operator is not unique. Although the $N_i$ commute with each other, in general the other generators of the sl(2)-triples do not. It is a non-trivial result of the sl(2)-orbit theorem \cite{CKS} that for each $N_i$ one can construct $n$ sets of commuting sl(2)-triples as\footnote{Here the lowering operator $N_i^-$ is closely related to $N_i$, but only $N_1^-=N_1$ holds generally.}
\beq
  \label{aht_001}
   \mbox{commuting sl(2)-triples:} \qquad \{N^-_i , N^+_i, N^0_i\}\,,
\eeq
with the standard commutation relations
\eq{
  \left[N_i^0 , N_j^{\pm} \right] = \pm 2 \op N_i^{\pm} \op \delta_{ij}\,,
  \hspace{50pt}
  \left[N_i^+ , N_j^{-} \right] = N_i^{0} \op \delta_{ij}\,.
}

\item The triples shown in \eqref{aht_001} can be used  to split
the vector space 
$H^4_{\rm prim}(Y_4, \mathbb R)$ 
in the following way (see appendix \ref{app:AHS} for technical details)
\eq{
  \label{decomp_001}
  H^4_{\rm prim}(Y_4, \mathbb R) = \bigoplus_{\lv\in \mathcal E} V_{\lv} \,,
  \hspace{50pt}
  \lv = (\ell_1, \ldots, \ell_{n})\,,
}
where $V_{\ell}$ are the eigenspaces of $N^0_i$ characterized by  $N_i^0 v_\lv=(\ell_i-\ell_{i-1}) v_\lv$, which can 
be rewritten as
\begin{equation}
(N^0_1+\ldots + N^0_i)\op v_\ell = \ell_i\op v_\ell\, , \qquad v_\ell \in V_\ell\, .
\end{equation}
The indices $\ell_i$ are integers that for Calabi-Yau four-folds  are in the range 
$\ell_i\in \{-4,\ldots, +4\}$,
and $\mathcal E$ denotes the set of all possible vectors $\lv$.

\item It is instructive to consider the periods of the (up to rescaling) unique, holomorphic $(4,0)$-form 
$\Omega$ of $Y_4$. Its period vector $\Pi$ admits an expansion in the strict asymptotic regime \eqref{sector_001} as
\begin{equation}\label{eq:pisl2}
\Pi_{\rm sl(2)} = e^{t^i N_i^-} \tilde{a}_0\, ,
\end{equation} 
where we did not display subleading polynomial terms of order $1/\gamma$ and exponentially suppressed corrections. We 
note that for derivatives of the periods both kinds of corrections can be essential, as it happens for example for the K\"ahler metric near a conifold point (see Appendix \ref{app:example}). These essential corrections are taken into account when computing e.g. the asymptotic form of the Hodge star operator. The leading polynomial term $\tilde{a}_0$ has a precise location in the sl(2)-decomposition
\begin{equation}
\mathop{\rm Re} \tilde{a}_0\, , \ \mathop{\rm Im} \tilde{a}_0 \in V_{d}\, ,
\end{equation}
with $d$ a vector of the form $d=(d_1, \ldots, d_n)$. The $d_i$ are the largest integers such that the condition $(N_1^-+\ldots +N_k^-)^{d_k} \tilde{a}_0 \neq 0$ is satisfied, i.e.~$\tilde{a}_0$ is a highest-weight state. We also note that $0\leq d_1 \leq \ldots \leq d_n \leq 4$.

\item We can make the above sl(2)-decomposition \eqref{decomp_001} more refined using 
 sl(2) highest-weight states and their descendants. Let us introduce the subspaces
 \begin{equation}\label{eq:primitive}
 P_\ell = V_\ell  \cap \ker \bigl[ (N_1^-)^{
\ell_1-\ell_0+1}\bigr] \cap \ldots \cap \ker \bigl[(N_n^-)^{
\ell_n-\ell_{n-1}+1}\bigr]\, ,
 \end{equation}
with $\ell_0\equiv 0$.  The decomposition \eqref{decomp_001} can then be rewritten as a 
weight-space decomposition in the following way
 \begin{equation}\label{eq:primitivedecomp}
  H^4_{\rm prim}(Y_4, \mathbb R) = \bigoplus_{\ell \in \cE_{\rm hw}}\,  \bigoplus_{k_1=0}^{\ell_1-\ell_0} \cdots \bigoplus_{k_n=0}^{\ell_n-\ell_{n-1}}  (N^-_1)^{k_1} \cdots (N^-_n)^{k_n} P_{\ell}\, ,
 \end{equation}
 where we defined the index set for highest-weight states as
 \begin{equation}\label{eq:Ehw}
 \cE_{\rm hw} = \{ \ell = (\ell_1 , \ldots, \ell_n)\ | \ 0 \leq \ell_1 \leq \ldots \leq \ell_n \leq 4 \}\, .
 \end{equation}
To summarize, the elements of $H^4_{\rm prim}(Y_4, \mathbb R)$ are arranged in irreducible representation of the 
boundary sl(2)-algebras. This means that all their information is encoded in the highest-weight subspaces $P_\ell$ (and in the lowering matrices), so that by successively applying $N_i^-$ the full primitive four-form cohomology  can be obtained.

We already note now that
by introducing a boundary Hodge decomposition, we will be able to break down the index set \eqref{eq:Ehw} into smaller components (c.f.~equation \eqref{Eprimexp}), thereby further reducing the set of allowed sl(2)-eigenspaces.
 
\end{itemize}


\subsection{Boundary Hodge decomposition}\label{sec:boundaryHodge}
In addition to the sl(2)-decomposition introduced above, there is another algebraic structure associated to the boundary:  the boundary Hodge decomposition given by 
\begin{equation}\label{eq:Hpqinfty}
H^4_{\rm prim}(Y_4,\mathbb{C}) = H^{4,0}_\infty \oplus H^{3,1}_\infty \oplus H^{2,2}_\infty \oplus H^{1,3}_\infty \oplus H^{0,4}_\infty\, ,
\end{equation}
 where $\overline{H^{p,q}_\infty}=H^{q,p}_\infty$.
This decomposition is independent of the limiting coordinates $t^i$. From a more physical perspective one can interpret \eqref{eq:Hpqinfty} as a charge decomposition, with the index $p$ of $H^{p,q}_\infty$ as charge, and its counterpart $q$ fixed by $p+q=4$.\footnote{This operator-based approach has been explored in more detail in \cite{Grimm:2020cda,Grimm:2021ikg, Grimm:2021idu}: for \eqref{eq:Hpqinfty} one can introduce a corresponding charge operator with $H^{p,q}_\infty$ as eigenspaces; for the sl(2)-triples one can rotate to a complex basis such that the generators commute with the charge operator. This thereby allows for a simultaneous decomposition into eigenstates, which -- although we do not use these operators explicitly in this work -- is described in this subsection.} We now discuss the following aspects:
\begin{itemize}
\item We first introduce a  Weil operator $\star_\infty$ that acts on an element in one of the subspaces $w_{p,q} \in H^{p,q}_\infty$ as
\begin{equation}
\star_\infty w_{p,q} = i^{p-q} w_{p,q}\, .
\end{equation}
The Weil operator thus squares to the identity, that is $\star_{\infty}^2 = \mathds 1$, and it is independent of the moduli $t^i$. It also maps the $V_{\lv} $ appearing in the sl(2)-decomposition \eqref{decomp_001} as
\eq{
  \label{cinf_001} 
  \star_{\infty} : V_{\lv} \to V_{-\lv} \,.
}
Here we have used that dim $V_{+\lv}=$ dim $V_{-\lv}$, which can be shown using the 
action of the sl(2)-triples. 

\item It is now instructive to split the highest-weight subspaces $P_\ell$ introduced in 
\eqref{eq:primitive}  as
\begin{equation}\label{eq:Pdecomp}
P_\ell = \bigoplus_{p+q=\ell_n+4} P_\ell^{p,q}\, ,
\end{equation}
based on the Deligne splittings associated to the strict asymptotic regime (see appendix \ref{app:AHS} for more details). For our purposes here it is sufficient to note that this splitting is correlated with the boundary Hodge decomposition given in \eqref{eq:Hpqinfty}, as becomes apparent by noting that
\begin{equation}\label{eq:actionCinf}
e^{iN^-_{(n)}} P_\ell^{p,q} \subseteq H^{p, 4-p}_\infty\, ,
\hspace{60pt}
p \geq q\, ,\qquad
N^-_{(n)}\equiv \sum_{i=1}^n N^-_i\,.
\end{equation}

\item Combining \eqref{eq:actionCinf} with \eqref{cinf_001}, expanding the exponentials and identifying sl(2)-eigenspaces, 
we can then show 
that
\begin{equation}\label{eq:actionCinf2}
 \star_\infty \left(\prod_{i=1}^n \frac{(iN^-_i)^{k_i}}{k_i!} v_\ell^{p,q} \right) = i^{2p-4}  \prod_{i=1}^n \frac{(iN^-_i)^{\ell_i-\ell_{i-1} -k_i}}{(\ell_i-\ell_{i-1} -k_i)!}  v_\ell^{p,q}\, ,
\end{equation}
for $v_\ell^{p,q}\in P^{p,q}_\ell$ (with $p\geq q$), and where $k_i$ are any set of integers that give a non-vanishing contribution on the left-hand side.\footnote{This identity has also appeared before in \cite{Bastian:2020egp} for the leading term $a_0$ of the periods, where it played an important role in determining charge-to-mass ratios of BPS states.} Let us note the similarity of this relation with the K\"ahler-form identity $\star J^k/k! = J^{d-k}/(d-k)!$ on a $d$-dimensional K\"ahler manifold. This identity will prove to be very useful in the study of moduli stabilization, where we need to identify self-dual fluxes under the Hodge star.

\item It is also helpful to correlate the highest-weights given in \eqref{eq:Ehw} with the charge decomposition described by \eqref{eq:Pdecomp}. This refined splitting of indices is given by
\begin{equation}\label{eq:Ehwcharge}
\cE_{\rm hw, charge} = \{ (p, q, \ell) \, | \ 0 \leq p,q \leq 4\, , \ \ell \in \cE_{\rm prim} , \ p+q=\ell_n +4 \} \,.
\end{equation}
For $p=4$ or $q=4$ there is one state -- $\tilde{a}_0$ obtained from \eqref{eq:pisl2} or its conjugate -- spanning the corresponding highest-weight subspace, which can be attributed to the Calabi-Yau condition $h^{4,0}=h^{0,4}=1$. The remaining highest-weight states have $1 \leq p,q \leq 3$ and weights similarly bounded as $0 \leq \ell_1 \leq \ldots \leq \ell_n \leq 2$. A more detailed explanation of this decomposition is given in appendix \ref{app:AHS}. Here we state its result for \eqref{eq:Ehwcharge} as
\begin{equation}\label{Eprimexp}
\cE_{\rm hw, charge} = \{ (4, d_n,  d), (d_n, 4, d) \} \cup \cE_{\rm K3}\, ,
\end{equation}
where we defined 
\begin{equation}\label{eq:EK3}
\cE_{\rm K3}= \{ (p,q, \ell)\,  | \ 1 \leq p,q \leq 3\, , \  0 \leq \ell_1 \leq \ldots \leq \ell_n \leq 2\, , \ p+q=\ell_n + 4\} .
\end{equation}
The remainder $\cE_{\rm K3}$ of highest-weight states are related to $(3,1)$-, $(2,2)$-, and $(1,3)$-forms in the boundary Hodge decomposition, as can be seen by using \eqref{eq:actionCinf}. Relabeling these $(p,q)$-forms by shifting both degrees down by one, we recover a decomposition reminiscent of the middle cohomology of a Calabi-Yau two-fold, namely a K3 surface. Likewise, the weights $\ell_i$ are bounded between zero and two. However, note that $(3,1)$-forms are not unique for four-folds, so this situation is analogous to having multiple copies of  K3-like blocks.

\item The decomposition \eqref{Eprimexp} is one of the key insights for our work on moduli stabilization later in this paper. It tells us that for only one or two sl(2)-representations the weights  lie within the range $-4 \leq \ell_i \leq 4 $ that characterizes four-folds, while the remainder of states have weights restricted as $-2 \leq \ell_i \leq 2 $. More concretely, we can decompose the primitive middle cohomology as
\begin{equation}\label{H4decompK3}
H^4_{\rm prim} ( Y_4, \mathbb{R}) = H_\Omega \oplus  \bigoplus_{\ell \in \cE_{\rm K3}} H_{{\rm K3},\ell} \, ,
\end{equation}
where we defined the subspaces
\begin{equation}\label{OmegaK3subspaces}
\begin{aligned}
H_{\Omega} &= \bigoplus_{k_1=0}^{d_1-d_0} \cdots \bigoplus_{k_n=0}^{d_n-d_{n-1}}  (N^-_1)^{k_1} \cdots (N^-_n)^{k_n} \big(P_{d}^{4,d_n} \oplus  P_{d}^{d_n,4}) \, , \\
H_{{\rm K3}, \ell} &= \bigoplus_{\substack{p,q \\
  (p,q , \ell) \in \cE_{\rm K3} }} \bigoplus_{k_1=0}^{\ell_1-\ell_0} \cdots \bigoplus_{k_n=0}^{\ell_n-\ell_{n-1}}  (N^-_1)^{k_1} \cdots (N^-_n)^{k_n} P^{p,q}_\ell\, .
\end{aligned}
\end{equation}
For the first equation in \eqref{OmegaK3subspaces} we take just the subspace $P^{4,4}_d$ once in the case that $d_n=4$. For the second equation the first sum means that, for a given set of highest weights $\ell$, we take all possible values $p,q$ such that $(p,q,\ell) \in \cE_{\rm K3}$. In \eqref{H4decompK3} we then sum over all blocks of K3 subspaces $H_{{\rm K3}, \ell}$ to recover the primitive cohomology $H^4_{\rm prim} ( Y_4, \mathbb{R}) $, where we suppressed the $p,q$ indices in the summation subscript.

For the purposes of moduli stabilization the majority of fluxes thus comes from the $H_{{\rm K3}, \ell}$: we will see that fluxes in $H_{\Omega}$ can only stabilize few moduli, so at large $h^{3,1}$ the essence of the problem is captured by the K3 subblocks. We stress that this splitting of the sl(2)-representations is a general result from asymptotic Hodge theory about the sl(2)-decomposition of $H^4_{\rm prim}(Y_4, \mathbb{R})$; the appearance of these blocks is not a simplifying assumption we make in this work, but a consequence of the existence of the boundary structure.

\end{itemize}


\subsection{The strict-asymptotic form of the Hodge star}
In this subsection we discuss the action of the Hodge-star operator $\star$ in the
strict asymptotic regime. For rigorous derivations of the relevant formulas we refer to \cite{CKS,Grimm:2018cpv}, while
here we only state that schematically we have
\beq
    \star\quad \xrightarrow{\quad \mbox{\scriptsize strict asymptotic regime}\quad } \quad \star_{\rm sl(2)} \,,
\eeq
where $\star_{\rm sl(2)}$ denotes the Hodge-star operator in the strict asymptotic regime. 
This operator can be written as the following matrix
\begin{equation}\label{eq:Csl2}
\star_{\rm sl(2)} = 
e^{+\phi^i N^-_i} 
\left[ e^{-\frac{1}{2} \log(s^i) N^0_i}\op \star_\infty \,e^{+\frac{1}{2} \log(s^i) N^0_i} \right]
e^{-\phi^i N^-_i}\, , 
\end{equation}
where $N_i^-$ and $N_i^0$ are elements of the sl(2)-triplets introduced in \eqref{aht_001}.
For later reference it is worthwhile to look more closely at the sl(2)-approximated Hodge decomposition:
\begin{itemize}
\item In analogy to \eqref{eq:Hpqinfty} we can write down a Hodge decomposition in the strict asymptotic regime as
\begin{equation} \label{Hodge sl2}
H^4_{\rm prim}(Y_4,\mathbb{C}) = H^{4,0}_{\rm sl(2)} \oplus H^{3,1}_{\rm sl(2)} \oplus H^{2,2}_{\rm sl(2)} \oplus H^{1,3}_{\rm sl(2)} \oplus H^{0,4}_{\rm sl(2)}\, ,
\end{equation}
with $\overline{H^{p,q}_{\rm sl(2)} } =H^{q,p}_{\rm sl(2)}$. As an example, the sl(2)-approximated period vector $\Pi_{\rm sl(2)}$ of the holomorphic $(4,0)$-form given in \eqref{eq:pisl2} spans the subspace $H^{4,0}_{\rm sl(2)}$. 

\item There is also a straightforward way to pass between the boundary Hodge structure $H^{p,q}_\infty$ and the sl(2)-approximated $H^{p,q}_{\rm sl(2)}$. We interpolate by applying the saxion- and axion-dependent factors in \eqref{eq:Csl2} as
\begin{equation}\label{eq:Hpqsl2}
H^{p,q}_{\rm sl(2)} = e^{\phi^i N^-_i} 
 e^{-\frac{1}{2} \log(s^i) N^0_i} H^{p,q}_{\infty}\, .
\end{equation}
This identity proves to be useful when lifting a boundary $(p,q)$-form to a $(p,q)$-form in the strict asymptotic regime. To be more explicit, by using \eqref{eq:actionCinf} for a highest-weight element $v^{p,q}_\ell \in P^{p,q}_\ell$ (with $p \geq q$) we can show with Baker-Campbell-Hausdorff that\footnote{In particular, one can prove that $e^{-\frac{1}{2} \log \left(s^{i}\right) N_{i}^{0}} \, e^{i N_{(n)}^{-}} \, e^{+\frac{1}{2} \log \left(s^{i}\right) N_{i}^{0}}\, =\, e^{i s^{i} N_{i}^{-}}$ in the strict asymptotic regime.}
\begin{equation}\label{eq:pqformsl2}
e^{i\op  t^i N_i^-} v^{p,q}_\ell \in H^{p,4-p}_{\rm sl(2)}\, .
\end{equation}
Note that the $(4,0)$-form period vector in \eqref{eq:pisl2} is a special case of this identity with $v^{4,d_n}_d = \tilde{a}_0$.

\end{itemize}
Our goal for the rest of this section is to determine the strict asymptotic limit of the norm defined in \eqref{norm&ip}.
We first note that for each of the subspaces $V_{\lv}$ appearing in the decomposition \eqref{decomp_001}, 
one can introduce a basis 
$\{(v_{\lv})_{ 1}, \ldots  (v_{\lv})_{ {\rm dim} V_{\lv}} \} \in V_{\lv}$. To find it, one can first introduce such a basis for the corresponding highest-weight subspaces, $P_\ell$ and then find the remaining basis elements by successive application of the $N_i^-$, in analogy with eq.~\eqref{eq:primitivedecomp}. This basis can then be normalized such that
\eq{
\label{eq:orthobasis}
  \langle \op (v_{\lv})_{i}, (v_{\lv'})_{j} \op\rangle = \pm \, \delta_{i, j}\op 
  \delta_{\lv, - \lv'} \,,
}
where $\langle \cdot,\cdot\rangle$ is the pairing defined in \eqref{norm&ip}, and depending on the particular vector within each $V_\ell$ the sign can be positive or negative.
We then define a norm $\lVert \op\cdot\op \rVert_{\infty}$ using the Weil operator $\star_{\infty}$  as follows
\eq{\label{eq:boundarynorm}
\lVert v_{\lv} \rVert^2_\infty=\langle \star_{\infty} v_{\lv}, v_{\lv} \rangle\,.
}
The pairing and norms for the basis elements associated with the K3-like blocks are explicitly shown in appendix \ref{app:Cinfty}. Next, we observe that the inner product of two vectors $v$ and $v'$ satisfies
\eq{
\label{innerprodinv}
\langle e^{\phi^i N^-_i } v , e^{\phi^j N^-_j } v' \rangle = \langle v, v' \rangle \,,
}
where 
$\phi^i$ are the axionic parts of the complex-structure moduli. 
Using  the explicit form of $\star_{\rm sl(2)}$ 
shown in 
\eqref{eq:Csl2} one can derive the following expression 
\eq{
\label{eq:growth}
 \left\langle  v_{\lv} ,  \left[ e^{-\phi^i N^-_i } \star_{\rm sl(2)} e^{+\phi^i N^-_i }\right] v_{\lv} \right\rangle
&=\left(s^1\right)^{\ell_1}\left(s^2\right)^{\ell_2-\ell_1} \ldots 
\left(s^n\right)^{\ell_n-\ell_{n-1}} \lVert v_{\lv}\rVert^2_\infty \\
&=   \left( \frac{s^1}{s^2} \right)^{\ell_1}   
\ldots \left( \frac{s^{n-1}}{s^{n}} \right)^{\ell_{n-1}}
  \left( s^{n}\right)^{\lv_{n}} \lVert  v_{\lv}\rVert^2_\infty 
\\[2pt]
&\equiv \kappa_{\lv}  \, \lVert  v_{\lv}\rVert^2_\infty \,.
}
We emphasize that $\lVert v_{\lv}\rVert^2_\infty$ is independent of the moduli and, explicitly, $\kappa_{\ell}$ is
given by
\begin{equation}
\label{kappaell}
  \kappa_{\lv} =   \left( \frac{s^1}{s^2} \right)^{\ell_1} \ldots \left( \frac{s^{n-1}}{s^{n}} \right)^{\ell_{n-1}}
  \left( s^{n}\right)^{\ell_{n}} \, . 
\end{equation}
Note that from this definition it is obvious that $\kappa_{-\lv}=(\kappa_{\lv})^{-1}$ where $-\ell$ stands for the vector with all entries opposite as those of $\ell$.


\subsubsection*{$V_{\text {heavy}}$, $V_{\text {light}}$ and $V_{\text {rest}}$}

Before closing this subsection, we want to define three subspaces of $H_{\rm{prim}}^{4}\left(Y_{4},\mathbb R\right)$
which are distinguished by the behavior of their norm in the strict asymptotic regime. 
We define $V_{\text {heavy}}$ and $V_{\text {light}}$
by
\begin{equation}
\arraycolsep2pt
\begin{array}{lcl@{\hspace{20pt}}ll}
\label{eq:Vheavylight}
V_{\text {heavy}} &= & \displaystyle \bigoplus_{\ell} V_{\ell} \, , &\text { with } \ell_1, \ell_2, \ldots , \ell_n \geq 0 & \text{and at least one }  \ell_i > 0 \, , 
\\[16pt]
V_{\text {light}} &= &\displaystyle  \bigoplus_{\ell} V_{\ell} \, , & \text { with } \ell_1, \ell_2, \ldots , \ell_n \leq 0 &\text{and at least one }   \ell_i < 0 \,,
\end{array}
\end{equation}
and $V_{\text{rest}}$ includes all the elements not belonging to $V_{\text {heavy}}$ nor $V_{\text {light}}$.
The  primitive  four-form cohomology splits as $H_{\text {prim}}^{4}\left(Y_{4}, \mathbb{R}\right)=V_{\text {heavy }} \oplus V_{\text {light }} \oplus V_{\text {rest }}$, and we come back to this splitting below.

 
\subsection{Explicit sl(2) subspaces on Calabi-Yau four-folds}
\label{ss:primitivesubspaces}

The expansion of $H_{\text{prim}}^{4}\left(Y_{4}, \mathbb{R}\right)$ into sl(2)-eigenspaces 
shown in \eqref{decomp_001} contains in general a large number of terms. 
However, in this section we  show that only a small number of sl(2)-subspaces  are populated, which makes possible a very explicit analysis of moduli stabilization.


\subsubsection*{Sl(2)-representations I}

In a first step we do not consider  the sl(2)-representation coming from the holomorphic $(4,0)$-form, 
i.e.~we focus on the subspaces corresponding to the second part in the set $\mathcal E$ 
shown in \eqref{Eprimexp}. 
(These  correspond to the inner part of the Deligne diamond \eqref{diamond}.)
According to  \eqref{Eprimexp} the possible highest weight states for this part are of the form 
\begin{equation}
\label{eq:highestweight}
P_0\,, \hspace{40pt} 
P_{0\, 1_i}\,, \hspace{40pt}
P_{0\, 2_i}\,, \hspace{40pt}
P_{0\, 1_i\, 2_j}   \,,
\end{equation}
where we employ the notation 
\eq{
  P_{0\, 1_i\, 2_j} =P_{0,\ldots,0,\underset{i}{1},\ldots,1,\underset{j}{2},\ldots,2}\,,
}
with $i$ and $j$ denoting the positions at which the first time a $1$ or $2$ appears. 
Note that this includes the possibility  $i=1$ for which a $0$ is absent, i.e.~the spaces $P$ do not need to start with a $0$.
All the sl(2)-representations can then be obtained by successively applying the lowering operators $N_i^-$, so that the full primitive four-form cohomology (up to the sl(2) state corresponding to the holomorphic $(4,0)$-form) can be decomposed into the following $V_{{\ell}}$ 
(repeated indices are not summed over)
\begin{equation}\label{eq:Vellspaces}
\begin{split}
\arraycolsep2pt
\begin{array}{lcl}
V_{\mathrm{heavy}}&=&\left\{ \begin{array}{lcl} 
V_{02_i}&=&P_{02_i}\\
V_{0 1_i}&=&P_{0 1_i}\\
V_{01_i2_j}&=&P_{01_i2_j}\\
V_{01_i 0_j}&=&N_j^- \, P_{01_i2_j}
\end{array}\right.
\\[34pt]
V_{\mathrm{light}}&=&\left\{ \begin{array}{lcl} 
V_{0-2_i}&=&(N_i^-)^2\, P_{02_i}\\
V_{0 -1_i}&=&N_i^- \, P_{0 1_i}\\
V_{0-1_i-2_j}&=& N_i^- \, N_j^- \, P_{01_i2_j}   \\
V_{0-1_i 0_j}&=&  N_i^- \, P_{01_i2_j}  
\end{array}\right.\, 
\end{array}
\hspace{30pt}
V_{\mathrm{rest}}=V_0=P_0 \,\oplus \, N_i^- \, P_{02_i} \,.
\end{split}
\end{equation}
We mention that a crucial point for what follows is the absence of $V_{\ell} \in \Vr $ with some positive 
and some negative $\ell_i$, that is,  $\Vr$ is entirely composed by $V_0$.


\subsubsection*{Sl(2)-representations II}

Let us now present the highest-weight states corresponding to the sl(2)-re\-pre\-sen\-ta\-tion coming from the holomorphic (4,0)-form. As will become clear in section~\ref{sec_modstab_explicitly}, these will not play an essential role in the stabilization of many moduli, but we display here for completeness. The possible highest-weight spaces are the following
\begin{equation}
\label{eq:highestweight40}
\arraycolsep10pt
\begin{array}{l l l l l l}
P_0 \, , & & & & & \\ 
P_{01_i} \, , &P_{02_i} \, , &P_{03_i} \, , &P_{04_i} \, ,  & & \\
P_{01_i 2_j}\, ,  &P_{01_i 3_j}\, , &P_{01_i 4_j}\, , &P_{02_i 3_j}\, , &P_{02_i 4_j}\, , &P_{03_i 4_j}\, ,  \\ 
P_{0\, 1_i\, 2_j 3_k} \, , &P_{0\, 1_i\, 2_j 4_k} \, , &P_{0\, 1_i\, 3_j 4_k} \, , &P_{0\, 2_i\, 3_j 4_k} \, ,  & & \\
P_{0\, 1_i\, 2_j 3_k 4_l} \, . & & & & &  
\end{array}
\end{equation}
Let us remark that, as opposed to the highest-weight states in equation \eqref{eq:highestweight}, for a given Calabi-Yau four-fold and for a given strict asymptotic regime (c.f.~eq.~\eqref{sector_001}) only one of these (4,0)-form highest-weight states will be present.\footnote{Note that by exploring different asymptotic limits for a given CY generally only a subset of the spaces shown in eq. \eqref{eq:highestweight40} (one for each asymptotic limit) are realized. That is, not necessarily all of them can be realized in a given CY.} This  corresponds to the outer part of the Deligne diamond \eqref{diamond}, and it can be directly related to the enhancement chain of the singularity types associated to the corresponding strict asymptotic regime.

By direct comparison of equations \eqref{eq:highestweight}  and \eqref{eq:highestweight40},
we see that the contribution from the (4,0)-form to the subspaces $\Vl$, $\Vh$ and $\Vr$ can be much richer than the one presented in \eqref{eq:Vellspaces}. We will not display all the possibilities here since, as mentioned, their detailed expression will not play a relevant role for our later discussion. They can however be obtained by applying all possible lowering operators to the corresponding highest-weight states.


\section{Moduli stabilization -- general considerations }
\label{sec_modstab_gen}

In this section we discuss complex-structure moduli-stabilization 
in the strict asymptotic limit using the framework of asymptotic Hodge
theory. We give an overview of the general structure, but 
provide a more detailed picture in section~\ref{sec_modstab_explicitly}.


\subsection{Self-duality condition}
\label{ss:self-duality}

Minkowski minima of the scalar potential \eqref{eq:potential} are obtained when solving 
the self-duality condition of the four-form flux $G_4$ shown in equation \eqref{cond_001}.
We now bring this condition into the framework of asymptotic Hodge theory. 
According to \eqref{decomp_001} we can expand $G_4$ as
$G_4 = \sum_{\lv\in\mathcal E} G_{\lv}$ with $G_{\lv} \in V_{\lv}$.
However, given the form of \eqref{eq:growth},  
it turns out to be  convenient to define an axion-dependent four-form flux 
and perform its expansion as \cite{Grimm:2019ixq}
\eq{
  \label{def_rho}
 \hG_4  \equiv e^{-\phi^i N^-_i } G_4 \,,\hspace{50pt}
  \hG_4 = \sum_{\lv \in \mathcal E} \hG_{\lv} \,,
}
where $\hG_{\lv} \in V_{\lv}$. Similar redefinitions using the log-monodromy matrices have been used in \cite{Bielleman:2015ina,Carta:2016ynn, Herraez:2018vae,Grimm:2019ixq,Marchesano:2019hfb,Grimm:2020ouv,Grimm:2021ckh}. Let us stress that in our setting the lowering operators $N_i^-$ appear, which in general are valued over the rationals instead of the integers. Note that an integral shift of the flux quanta can therefore require the axions to wind around multiple times instead of just once. The self-duality condition  \eqref{cond_001} in the strict asymptotic 
regime  reads $G_4 = \star_{\rm sl(2)} G_4$, which can be  rewritten as
\eq{
  \label{exp_735691}
  \hat G_4 = \left[ e^{-\phi^i N^-_i } \star_{\rm sl(2)} e^{+\phi^i N^-_i }\right]  \hat G_4  \,.
}
Comparing now with \eqref{eq:Csl2} and recalling that the $N_i^0$ 
act as $N_i^0: V_{\lv}\to V_{\lv}$ while $\star_{\infty}: V_{\lv}\to V_{-\lv}$, 
we note that the expression in parenthesis in \eqref{exp_735691}
maps $V_{\lv}$ to $V_{-\lv}$. 
Taking the inner product with $\hat G_{\lv}$ and 
using the orthogonality condition \eqref{eq:orthobasis} 
as well as \eqref{eq:growth} we obtain
\eq{
  \label{mod_stab_102}
 \langle \hat G_{+\lv} , \hat G_{-\lv} \rangle = 
 \left \langle \hat G_{\lv} ,
 \left[ e^{-\phi^i N^-_i } \star_{\rm sl(2)} e^{+\phi^i N^-_i }\right]  \hat G_{\lv} \right\rangle
 = \kappa_{\lv} \op \lVert \hat G_{\lv} \rVert_{\infty}^2\,.
}
Let us emphasize that the axions  appear only in $\hat G_{\lv}$ 
and  the saxions appear only in $\kappa_{\lv}$. 
We furthermore note that the norm $\lVert \,\cdot\, \rVert_{\infty}$ is positive-definite and 
that in our conventions the saxions  satisfy $s^{i}>0$.
Hence, \eqref{mod_stab_102} implies the two relations
\begin{align}
  \label{mod_stab_202}
  &\langle \hG_{+\lv} , \hG_{-\lv} \rangle \geq 0 \,,
  \\[4pt]
  \label{mod_stab_202b}
   &\hG_{+\lv}\neq0 \quad\Rightarrow\quad \hG_{-\lv}\neq0 \,.
\end{align}   
We also observe that the pairing $\langle \hG_{+\lv} , \hG_{-\lv} \rangle$ is invariant under 
$\lv \to -\lv$, and so we obtain from \eqref{mod_stab_102} the relation
$ \langle \hat G_{+\lv} , \hat G_{-\lv} \rangle =  \kappa_{-\lv} \op \lVert \hat G_{-\lv} \rVert_{\infty}^2$.
Multiplying this expression with \eqref{mod_stab_102}
leads to 
\eq{
  \label{relation_001}
  \langle \hG_{+\lv} , \hG_{-\lv} \rangle^2 =  
  \lVert  \hG_{+\lv} \rVert_{\infty}^2 \;
  \lVert  \hG_{-\lv} \rVert_{\infty}^2 \,,
}
where we used that $\kappa_{\lv}\, \kappa_{-\lv} = 1$. 
This relation corresponds to the equality in a Cauchy-Schwarz inequality for the 
pairing
$\langle \,\star_{\infty} \, \cdot\,, \,\cdot\, \rangle$, which can only be satisfied for 
\eq{
  \label{relation_101}
  \hG_{-\lv} = \kappa_{\lv} \,  \star_{\infty} \op \hG_{+\lv} 
  \, . 
}


\subsubsection*{The tadpole}

Let us now turn to the tadpole contribution of the flux $G_4$ given in \eqref{charge_q}.
In terms of sl(2)-representations this is
\eq{
  \label{tadpolel}
 Q = \frac{1}{2}\op\langle G_4 , G_4 \rangle = 
 \frac{1}{2}\op\langle \hG_4 , \hG_4 \rangle = 
 \frac{1}{2}\sum_{{\lv}} \langle \hG_{+\lv}, \hG_{{-\lv}} \rangle  \,,
 }
where we have used  \eqref{relation_001}. We can use the self-duality condition, and the properties of the different sl(2)-weights to give a more explicit expression, as we will see in section~\ref{sec:tadpolecontribution}.


\subsection{Saxion stabilization}
\label{ss:saxionstab}

Let us now discuss the stabilization of the saxions $s^i$ through the conditions \eqref{relation_101}.
(We address the stabilization of the axions in section~\ref{ss:axionstab} below.) 
Taking the ratio between equation \eqref{mod_stab_102} and its version with $\ell \rightarrow -\ell$, and 
using $\langle\hat{G}_{+\ell}, \hat{G}_{-\ell}\rangle=\langle\hat{G}_{-\ell}, \hat{G}_{+\ell}\rangle$ and
$\kappa_{-\lv}= \kappa_{+\lv}^{-1}$,
we arrive at 
\begin{equation}
\label{eq:ratiorhos}
\kappa_{\lv}=\dfrac{ \lVert  \hG_{-\ell} \rVert_{\infty}}{ \lVert  \hG_{+\ell} \rVert_{\infty}} 
\hspace{50pt}\mbox{with}\qquad
\kappa_{\ell}=\left(\frac{s^{1}}{s^{2}}\right)^{\ell_{1}} \cdots\left(\frac{s^{n-1}}{s^{n}}\right)^{\ell_{n-1}}\left(s^{n}\right)^{\ell_{n}}\,.
\end{equation}
As we have seen in \eqref{mod_stab_202b}, if $\hat G_{+\lv}\neq0 $ then also 
$\hat G_{-\lv}\neq0 $. Let us therefore label all non-vanishing $\hG_{\lv}$ by an index $\alpha$
and define
\eq{
  \label{mod_stab_105}
  y^{i} = \log \frac{s^{i}}{s^{i+1}} \,, \hspace{60pt} 
  \mathcal B_{\alpha} = \log \dfrac{ \lVert  \hG_{-\alpha} \rVert_{\infty}}{ \lVert  \hG_{+\alpha} \rVert_{\infty}}
  \,,
}
together with $s^{n+1}\equiv1$. Note that in the strict asymptotic limit we are considering 
(c.f.~equation \eqref{sector_001}), all the $y^i$ are positive. 
Taking then the logarithm of equation \eqref{eq:ratiorhos} leads to
\eq{
  \label{mod_stab_106}
  \ell_{(\alpha)1}\, y^1 + \ell_{(\alpha)2} \,y^2 + \ldots + \ell_{(\alpha)n} \,y^{n} = \mathcal B_{\alpha} \,,
}
which in matrix notation is expressed as
\eq{
  \label{sdc_001}
  \mathcal A_{\alpha i}\, y^i = \mathcal B_{\alpha} \,.
}
The number of saxions $s^i$ stabilized by this condition is equal to the rank of $\mathcal A$,
and in order to stabilize all saxions we have to require $\mathcal A$ to 
be of maximal rank, that is ${\rm rank}\, \mathcal A = h^{3,1}$. \label{page_saxion_stab}
This implies in particular, that we need to have at least $h^{3,1}$ 
non-vanishing pairs $(  \hG_{+\lv} ,  \hG_{{-\lv}})$.
The values of the stabilized saxions $s^i$ are then determined via the relation
\eq{
  \label{sdc_003}
  y^i = \left[ \mathcal A^{+}\right]^{i \beta} \mathcal B_{\beta} \,,
}
where $\mathcal A^+$ denotes the pseudo-inverse of $\mathcal A$ which can be computed for instance
through a singular value decomposition.  
Note that when ${\rm rank}\, \mathcal A = h^{3,1}$ and $\mathcal B=0$ we have $y^i=0$, which 
is not compatible with the growth requirement \eqref{sector_001}. We therefore require at least one 
$\kappa_{\alpha}\neq 1$.


\subsection{Axion stabilization}
\label{ss:axionstab}

Let us now briefly consider the stabilization of the axions $\phi^i$ from a general perspective,
while 
a more detailed discussion of this question will be given in section~\ref{sec_modstab_explicitly}
below. 
We first recall that the axions $\phi^i$ appear in the self-duality condition \eqref{relation_101} 
through
the relation \eqref{def_rho}, that is
\eq{
  \label{rel_38465}
 \hG_4  = e^{-\phi^i N^-_i } G_4 \,,
}
where $N^-_i$ are the lowering operators  in the 
commuting sl(2)-triples \eqref{aht_001}.
The flux $G_4$ can be expanded into the sl(2)-eigenspaces $V_{\lv}$ 
appearing in \eqref{decomp_001} as
\eq{
  G_4 = \sum_{\lv\in\mathcal E} G_{\lv} \,.
}
Now, if $G_4$ has only components $G_{\lv}$ which are all annihilated by 
the  action of a particular $N_{ i}^-$, then the corresponding axion $\phi^{ i}$ 
does not appear in \eqref{rel_38465} and will not be stabilized. 
Generalizing this argument, we therefore have at most
\eq{
  \mbox{dim}\,\mbox{span}  \{ N_i^- G_{4} \}
}
linearly independent combinations of axions $\phi^i$ appearing in $\hG_4$, which is therefore the maximal number of axions that can be stabilized
through the self-duality condition \eqref{relation_101}.


\section{Moduli stabilization -- explicit analysis}
\label{sec_modstab_explicitly}

After having discussed moduli stabilization from a general point of view in section~\ref{sec_modstab_gen},
we now turn to a more detailed treatment. We 
are going to make use of the decomposition of 
$H_{\text{prim}}^{4}\left(Y_{4}, \mathbb{R}\right)$ into $V_{\text {heavy }}$, $V_{\text {light }}$ and $V_{\text {rest }}$
shown in section~\ref{ss:primitivesubspaces}. Let us note that we will mostly ignore the sl(2)-eigenspaces coming from the irreducible representation(s) of the $(4,0)$-form and focus our attention on those coming from the middle part.
As explained below, the former do not play an important role for  large numbers of moduli.


\subsubsection*{Decomposition using K3-like blocks} 

Recall from section \ref{ss:primitivesubspaces} that the highest-weight subspaces in the sl(2)-de\-com\-po\-sition \eqref{eq:primitivedecomp} can be divided into a one- or two-dimensional part corresponding to the $(4,0)$-form (and its conjugate) with weights $0\leq d_1 \leq \ldots \leq d_n \leq 4$. For all others these are bounded by $0\leq \ell_1 \leq \ldots \leq \ell_n \leq 2$. To be more explicit, for completeness we restate \eqref{H4decompK3} as
\begin{equation}
H^4_{\rm prim} ( Y_4, \mathbb{R}) = H_{\Omega} \oplus  \bigoplus_{\ell \in \cE_{\rm K3}} H_{{\rm K3},\ell} \, ,
\end{equation}
where the subspaces on the right-hand side were defined in \eqref{OmegaK3subspaces}. The first piece $H_{\Omega} $ denotes the sl(2)-eigenspaces that descend from the highest-weight state of the $(4,0)$-form and its complex conjugate. The remainder $H_{{\rm K3},\ell} $ (with $\ell \in \cE_{\rm K3}$) is comprised of descendants of highest-weight states corresponding to $(3,1)$-, $(2,2)$- and $(1,3)$-forms. For the purposes of moduli stabilization we can stabilize at most four moduli (eight real scalars) via fluxes in $H_{\Omega} $, since this is the maximum  number of  different $V_\ell$ that one can populate in this sl(2)-representation. For any strict asymptotic regime with large $h^{3,1}$ the majority of moduli thus has to be stabilized through fluxes in subspaces $H_{{\rm K3},\ell}$. In the following we will therefore restrict our attention to fluxes in $H_{{\rm K3},\ell}$, and comment afterwards on the inclusion of fluxes in $H_{\Omega}$.


\subsubsection*{Hodge decomposition of K3-like blocks} 

For our study of moduli stabilization it is useful to write out a Hodge decomposition for $H_{{\rm K3},\ell}$ in the strict asymptotic regime as
\begin{equation}
H_{{\rm K3},\ell} = (H_{{\rm K3},\ell})_{\rm sl(2)}^{3,1} \oplus  (H_{{\rm K3},\ell})_{\rm sl(2)}^{2,2} \oplus (H_{{\rm K3},\ell})_{\rm sl(2)}^{1,3}\, .
\end{equation}
For the self-duality condition on the fluxes \eqref{cond_001} we can only allow for $G_4 \in (H_{{\rm K3},\ell})_{\rm sl(2)}^{2,2} $. In particular, we cannot have $(4,0)$- or $(0,4)$-fluxes coming from these subspaces, so the vacuum loci will be supersymmetric by construction, i.e.~as many axions and saxions will be stabilized, both with the same mass.\footnote{One can also include supersymmetry-breaking pieces coming from fluxes in $H_{\Omega}$, as discussed at the end of section \ref{sec:vacuawithVK3}. These on one hand do not alter the analysis of the vacuum loci, and on the other they can only fix a handful of moduli, so we do not consider them here.} 
As a complementary perspective, it is convenient to rewrite the self-duality condition restricted to $H_{{\rm K3},\ell}$ into an orthogonality condition with $(H_{{\rm K3},\ell})_{\rm sl(2)}^{3,1} $. Recalling \eqref{eq:pqformsl2}, we can write down a basis for these vector spaces in terms of the highest-weight states \eqref{eq:Pdecomp} as 
\begin{equation}
 \chi_\ell(t^i) = e^{i t^i N^-_i} v^{3,1+\ell_n}_\ell \in (H_{{\rm K3},\ell})_{\rm sl(2)}^{3,1} \, , \hspace{50pt} v^{3,1+\ell_n}_\ell \in P^{3,1+\ell_n}_\ell \,.
\end{equation}
The self-duality condition restricted to $H_{{\rm K3},\ell} $ is then implemented by
\begin{equation}\label{eq:H31ortho}
\langle e^{i t^i N^-_i} v^{3,1+\ell_n}_\ell , G_4 \rangle = 0\, .
\end{equation}
This gives us a simple set of algebraic equations, holomorphic in the moduli, that we need to solve to determine the vacuum loci for a given flux $G_4 \in H_{{\rm K3},\ell} $.


\subsection{No moduli stabilization with fluxes in $\Vr \subset H_{K3}$}
\label{nomodstabVrest}
We start by considering fluxes in $\Vr$. 
As can be seen from equation \eqref{eq:Vellspaces}, $\Vr$ splits into 
$P_0$ and $N_i^- \, P_{02_i} $ which we discuss in turn. 
First,  $P_0$ is the sl(2)-singlet that is annihilated by all 
$N^-_i$, which implies for  $G_4 \in P_0$ that 
 \beq
G_4  \in P_0 \hspace{30pt}\Rightarrow\hspace{30pt} G_4=\hat G_4 \,.
 \eeq
The self-duality condition \eqref{relation_101} then reads
\beq
 G_4=\star_{\infty}\op G_4 \,,
\eeq
in particular, it
is independent of the axions and saxions. Hence, through such fluxes no moduli 
are stabilized. Moreover, the only fluxes in $P_0$ that are self-dual, namely the ones in $P_0^{(2,2)}$, do not contribute to the potential (see eq.~\eqref{eq:pairingCinftyP022}). 
Second, fluxes in $\Vr$ that are not highest-weight, namely $G_4 \in N_i^- \, P_{02_i} $, are anti-self dual (see equation \eqref{eq:CinftyP02}) and therefore do not satisfy the
self-duality condition \eqref{relation_101}.
To summarize, fluxes in $\Vr$ do not stabilize moduli. 

Let us note that the situation for the fluxes in $\Vr$ coming from the (4,0)-form can be slightly more involved, and such fluxes can indeed be used to fix moduli. However, as emphasized at the beginning of this section, this will not play an important role when trying to fix many of moduli, since at most four (complex) moduli can be fixed by turning on these fluxes.


\subsection{Vacua with fluxes in $\Vh$ and $\Vl$}
\label{sec:vacuawithVK3}
Let us consider the moduli stabilization including fluxes in $\Vh$, given by the spaces in \eqref{eq:Vellspaces}. The self-duality condition will relate these fluxes to their counterparts in $\Vl$, so we will also introduce these. We will consider the situation where there is flux in each of these individual subspaces, and then argue that combining them leads to the same conclusions.


\subsubsection*{Flux along $V_{01_i}$}

We begin by considering a highest-weight four-form flux $G_{01_i}^{R} v_{01_i}^{R}+G_{01_i}^I v_{01_i}^{I}\in V_{01_i}=P_{01_i}$, where we denote by $G_{01_i}^{R,I}$  the two possible flux component along the  $V_{01_i}$ space in the basis introduced in eq.~\eqref{eq:basisP01}. Note that each $V_{01_i}$ has (real) dimension 2, so that we need to allow for both kinds of fluxes. The flux-axion polynomial then reads
\begin{equation}
\hG_4= G_{01_i}^{R} v_{01_i}^{R}+G_{01_i}^I v_{01_i}^{I}+ \left( G_{0-1_i}^R-\phi^i  G_{01_i }^R\right)v_{0-1_i}^R + \left( G_{0-1_i}^I-\phi^i  G_{01_i }^I\right)v_{0-1_i}^I \, ,
\end{equation}
where no sum over $i$ is implied and where we have explicitly substituted the basis elements shown in eq.~\eqref{eq:basisP01}. With this flux configuration, the tadpole \eqref{tadpolel} is ${Q = G_{01_i}^R G_{0-1_i}^I-G_{01_i}^I G_{0-1_i}^R\, } $,  and the self-duality conditions read
\begin{equation}
\begin{split}
s^i  G_{01_i}^R \, &=\,  G_{0-1_i}^{I}-\phi^i  G_{01_i }^I \, , \\
-s^i  G_{01_i}^I \, &=\, G_{0-1_i}^{R}-\phi^i  G_{01_i }^R \, .
\end{split}
\end{equation}
These equations have solutions whenever at least one of the two pairs of fluxes $(G_{01_i}^R, G_{0-1_i}^I)$ or $(G_{01_i}^I, G_{0-1_i}^R)$ are non-zero. 
For concreteness, if one considers for instance the former, the moduli are fixed as\footnote{If the other pair is considered, a similar solution is obtained upon exchanging $(G_{01_i}^R, G_{0-1_i}^I) \rightarrow (-G_{01_i}^I,G_{0-1_i}^R)$. If any of such pairs is turned on and also the component of the other pair along $\Vl$, the result is just a shift in the vev of the axion $\phi^i \rightarrow \phi^i +G_{0-1_i}^I/G_{01_i}^R$ in the solution in \eqref{eq:minP01simple}. } 
\begin{equation}
\label{eq:minP01simple}
s^i =\dfrac{G_{0-1_i}^I}{G_{01_i}^R}\, , \qquad \phi^i=0\,, 
\end{equation}
The sign of $s^i$ is the same as the sign of the tadpole, so that a physical solution (i.e. $s^i >0$) requires a positive contribution to the tadpole. If both pairs of fluxes are turned on there is also a solution provided that the tadpole is positive, that is
\begin{equation}
\label{eq:minP01general}
s^i =\dfrac{G_{01_i}^R G_{0-1_i}^I-G_{01_i}^I G_{0-1_i}^R}{(G_{01_i}^I)^2(G_{01_i}^R)^2}\, , \qquad \phi^i=\dfrac{G_{01_i}^I G_{0-1_i}^I+G_{01_i}^R G_{0-1_i}^R}{(G_{01_i}^I)^2(G_{01_i}^R)^2}\, , 
\end{equation} 
and the tadpole appears in the numerator of the saxion vev. 


\subsubsection*{Flux along $V_{02_i}$}
We consider now the four-form highest-weight flux in $V_{02_i}=P_{02_i}$ and its descendants, and use the basis presented in eq.~\eqref{eq:basisP02}. The flux-axion polynomial is now
\begin{equation}
\hG_4= G_{02_i} v_{02_i} + \left( G_{0} -\phi^i  G_{02_i }\right) v_0+\left( G_{0-2_i} -\phi^i  G_{0}+\dfrac{1}{2}(\phi^i)^2 G_{02_i} \right)v_{0-2_i} \, ,
\end{equation}
and the corresponding tadpole reads ${Q= G_{02_i} G_{0-2_i}-\frac{1}{2} G_0^2}$. The self-duality condition gives the following equations 
\eq{(G_0 -\phi^i G_{02_i}) &= -(G_0 -\phi^i G_{02_i}) \quad \Longrightarrow  \quad G_0 -\phi^i G_{02_i}=0 \,,  \\  
\dfrac{G_{02_i}}{2} (s^i)^2   &=   G_{0-2_i} -\phi^i  G_{0}+\dfrac{1}{2}(\phi^i)^2 G_{02_i} \, .
}
Note that moduli stabilization requires $G_{02_i} \neq 0$. 
Without  loss of generality, we can use the axionic shift symmetry to set $G_0=0$ and we get the solution 
\begin{equation}
\label{eq:minP02simple}
s^i=\sqrt{\dfrac{2G_{0-2_i}}{G_{02_i}}}\, ,  \qquad \phi^i =0 \, ,
\end{equation}
 where again a real positive saxion implies a positive tadpole. 
Note that the general solution can be obtained from  \eqref{eq:minP02simple} by shifting in the axionic field $\phi^i \rightarrow \phi^i + \phi^i_b$, together with  $G_4 \rightarrow e^{\phi^i_b N_i^-}G_4$. This yields the general solution
\begin{equation}
\label{eq:minP02general}
s^i=\dfrac{\sqrt{2 \, G_{0-2_i} G_{02_i} - G_0^2}}{G_{02_i}}\, ,  \qquad \phi^i=\dfrac{G_0}{G_{0-2_i}}\, .
\end{equation}


\subsubsection*{Flux along $V_{01_i2_j}$}

Let us finally consider the highest weight flux in $V_{01_i2_j}$, together with its descendants. Using the basis in  eq.~\eqref{eq:basisP012}, the corresponding flux-axion polynomial reads
\eq{\label{eq:G4V012}
&\hG_4 =  G_{01_i2_j} v_{01_i2_j} + \left( G_{01_i0_j}-\phi^j G_{01_i2_j}\right)v_{01_i0_j}+ \left( G_{0-1_i0_j}-\phi^i G_{01_i2_j}\right) v_{0-1_i0_j} \\[4pt]
&\hspace{27pt}
+\left( G_{0-1_i-2_j}-\phi^i G_{01_i0_j}-\phi^j G_{0-1_i0_j}+ \phi^i \phi^j G_{01_i0_j} \right) v_{0-1_i-2_j}\, .
}
The tadpole contribution of these fluxes is ${Q= G_{01_i2_j}G_{0-1_i-2_j}- G_{01_i0_j}G_{0-1_i0_j}}$. Using the action of $\star_\infty$ on the different basis vectors, displayed in \eqref{eq:CinftyP012}, the self-duality condition yields the following two equations.
\eq{\label{selfduality012}
&s^i \, s^j\,  G_{01_i2_j} = \left( G_{0-1_i-2_j}-\phi^i G_{01_i0_j}-\phi^j G_{0-1_i0_j}+ \phi^i \phi^j G_{01_i0_j} \right) , \\
&-\dfrac{s^i}{s_j}  \left( G_{01_i0_j}-\phi^j G_{01_i2_j}\right) = \left( G_{0-1_i0_j}-\phi^i G_{01_i2_j}\right) \, .
}
Instead of solving these equations directly, it is more practical to switch to the orthogonality condition \eqref{eq:H31ortho} with the sl(2)-approximated $(3,1)$-form. We find that the $(3,1)$-form is given by
\begin{equation}
\begin{aligned}
\chi_{01_i2_j} &= \big(1+t^i N_i^-\big) \big(1 + t^j N_j^- \big) v_{01_i2_j}\\
&=  v_{01_i2_j} +t^i  v_{0-1_i0_j}  +t^j v_{01_i0_j} +t^i t^j v_{0-1_i-2_j} \, .
\end{aligned}
\end{equation}
When $G_{01_i2_j}= 0$, the equations simplify considerably and we can use the shift symmetry of the axions to set $G_{01_i2j}$ to zero without loss of generality. The solution then reads 
\begin{equation}
\label{eq:minP012_1simple}
s^j=-\dfrac{G_{01_i0_j}}{G_{0-1_i0_j}} s^i\,  ,\qquad \phi^j= -\dfrac{G_{01_i0_j}}{G_{0-1_i0_j}} \phi^i\, ,
\end{equation} 
which has physical solutions with positive saxions only when the tadpole is positive. 

When $G_{01_i2_j}\neq 0$, requiring orthogonality of $G_4$ with $\chi_{01_i2_j}$ then yields\footnote{Up to proportionality factors, the real part of this equation corresponds to the first equation in \eqref{selfduality012}, and the imaginary part to the second.}
\begin{equation}
(G_{01_i 2_j} t^i - G_{0-1_i0_j} )(G_{01_i 2_j}t^j - G_{01_i 0_j} ) = G_{01_i0_j} G_{0-1_i0_j}-G_{01_i2_j}G_{0-1_i-2_j}\, ,
\end{equation}
which can easily be solved for $t^i=\phi^i+i s^i$ or $t^j=\phi^j+i s^j$ by moving the other factor to the right-hand side. For concreteness, let us write the explicit solution for the case $G_{01_i0_j}=G_{0-1_i0_j}=0$ (from which the general case  above can also be obtained by exploiting the shift-symmetries of the axions):
\begin{equation}
\label{eq:minP012_2simple}
s^j\, =\, \dfrac{G_{0-1_i-2_j}}{G_{01_i2_j}}\ \dfrac{s^i}{(s^i)^2 + (\phi^i)^2}\,  , \qquad \phi^j\, =\, -\dfrac{G_{0-1_i-2_j}}{G_{01_i2_j}}\ \dfrac{\phi^i}{(s^i)^2 + (\phi^i)^2} \, .
\end{equation}
As before, this is a physical solution with positive saxions only when the signs of the relevant fluxes are such that the tadpole is positive.
 Thus, one positive contribution to the tadpole is able to fix one linear combination of the saxions and of the axions.


\subsubsection*{Flux along all K3-like subspaces}

Let us  discuss the situation with fluxes along an arbitrary combination of subspaces in 
 $\Vh$ and $\Vl$. Since self-duality conditions for each subspace decouple, each of the individual equations will have to be satisfied. This means that there will generically be no solutions, unless the fluxes satisfy certain properties. For instance, for fluxes in $V_{01_i}$ and $V_{02_i}$ at the same time, there is a solution only if \eqref{eq:minP01simple} and \eqref{eq:minP02simple} are satisfied, which implies a relation between the ratios of fluxes appearing in these equations. On the other hand, these pairs contribute to the tadpole, so tadpole-wise the most economic way of fixing the modulus $t^i$ moduli is having a single pair of fluxes either of the type $V_{01_i}$ or of the type $V_{02_i}$. In order to fix two moduli $t^i$ and $t^j$, one can either use $V_{01_i}$ or $V_{02_i}$ together with $V_{01_j}$ or $V_{02_j}$, or alternatively any of these four together with $V_{01_i2_j}$. Any of these possibilities will fix both moduli in the most economic way tadpole-wise. We come back to this point in section \ref{sec:tadpolecontribution}.


\subsubsection*{Adding fluxes along $H_\Omega$}

Finally, let us comment on the effect of including the fluxes in $H_\Omega$ coming from the sl(2)-representation whose highest-weight state corresponds to the (4,0)-form (and its complex conjugate). We do not perform a systematic study of moduli stabilization with all possible flux choices in $H_\Omega$ here, but we explain instead why it is not relevant for determining the scaling of the tadpole charge with a large number moduli stabilized in the strict asymptotic region.

Using the expression \eqref{eq:Csl2} for the Hodge star it can be seen that at most $d_n \leq 4$ moduli can appear in the self-duality condition for such fluxes. Thus, a large number of moduli cannot be stabilized by means of such fluxes, as opposed to the ones in $\bigoplus_{\ell \in \cE_{\rm K3}} H_{{\rm K3},\ell}$. Moreover, the fact that the self-duality condition for each subspace decouples implies that if these fluxes yield equations for some moduli that also appear in the conditions coming from the K3-like fluxes, they will only be satisfied for compatible flux configurations, but they will never lower the tadpole. As before, tadpole-wise the most economic way to fix moduli is by choosing fluxes in such a way that the moduli that appear in the conditions for the fluxes in $H_\Omega$ do not appear in the rest. This will still lead to (at least) a linear scaling of the tadpole with the (large) number of stabilized moduli, coming from the K3-like fluxes.

It is also interesting to consider the effect of supersymmetry breaking introduced by these fluxes. As mentioned at the beginning of the section, the K3-like fluxes do not induce any supersymmetry breaking effect, so that fixing moduli using only them always yields supersymmetric vacua, where both the saxions and their corresponding axions are fixed and have equal mass. 
However, this does not mean that our analysis cannot capture non-supersymmetric vacua. This is the case due to the decoupling of the self-duality conditions for the different subspaces, which crucially ensures that the inclusion of fluxes in $H_\Omega$ does not modify the scaling of the tadpole  with the number of fixed moduli.  Note, however,  that these supersymmetry breaking fluxes can alter the masses of the moduli, as they can introduce some mixing at the level of the scalar potential, even though they cannot do it at the level of the vacuum equations. To sum up, adding fluxes in $H_\Omega$ does not change the discussion with respect to the vacuum loci or the scaling of the tadpole with the number of fixed moduli, but it can change the values of the masses of the moduli, as expected from the fact that they can break supersymmetry.


\section{The tadpole contribution}
\label{sec:tadpolecontribution}

In this section we analyze the scaling of the flux-induced tadpole charge $Q$ with the number of fixed 
moduli. We compare this with the behavior predicted by the tadpole conjecture \cite{Bena:2020xrh}, 
which was  introduced in section \ref{ss:tadpoleconjecture}. 


\subsubsection*{Linear scaling with $h^{3,1}$}

Let us start by recalling the tadpole contribution of the flux $G_4$ from equation \eqref{charge_q} 
as $Q = \frac{1}{2} \int G_4\wedge G_4= \frac{1}{2} \langle G_4, G_4\rangle$. 
Using the self-duality condition \eqref{relation_101},  the 
property \eqref{innerprodinv} and the condition \eqref{relation_001} we compute
\eq{
  \label{tadpole_044b}
 \langle G_4 , G_4 \rangle = \langle \hG_4 , \hG_4 \rangle = \sum_{{\lv}} \langle \hG_{+\lv}, \hG_{{-\lv}} \rangle  
 = \sum_{{\lv}} \, \lVert  \hG_{\lv} \rVert_{\infty}\,
  \lVert  \hG_{-\lv} \rVert_{\infty} \,.
}
As argued below equation \eqref{sdc_001}, if we want to stabilize $n_{\rm stab}$ saxions  we need to have $n_{\rm stab}$ non-vanishing
pairs $(  \hG_{+\lv} ,  \hG_{{-\lv}})$ and hence the sum \eqref{tadpole_044b} 
contains at least $2\op n_{\rm stab}$ non-vanishing positive terms. As shown explicitly in the previous section, this would also stabilize   the corresponding $n_{\rm stab}$ axions. 
Using then for instance \eqref{eq:ratiorhos}, we can rewrite \eqref{tadpole_044b} in the following way
\eq{
  \label{tadpole_044c}
 \langle G_4 , G_4 \rangle \hspace{2pt}
  \arraycolsep2pt
  \begin{array}[t]{ccc@{}lcc@{}lcc@{}l}
  = && \displaystyle \sum_{{\lv}} &
   \kappa_{\lv} \lVert  \hG_{\ell} \rVert_{\infty}^2  
   \\[14pt]
   =&&\displaystyle \sum_{{\lv}\in V_{\rm heavy}} &
   \kappa_{\lv} \lVert  \hG_{\ell} \rVert_{\infty}^2 &+& \displaystyle\sum_{{\lv}\in V_{\rm rest} } &
  \kappa_{\lv} \lVert  \hG_{\ell} \rVert_{\infty}^2 &+&\displaystyle \sum_{{\lv} \in V_{\rm light}}  &
   \kappa_{\lv} \lVert  \hG_{\ell} \rVert_{\infty}^2 
   \\[16pt]
   =&2&\displaystyle \sum_{{\lv}\in V_{\rm heavy}} &
   \kappa_{\lv} \lVert  \hG_{\ell} \rVert_{\infty}^2  &+&  \displaystyle\sum_{{\lv} \in V_{\rm rest}} &
  \kappa_\ell \lVert  \hG_{\ell} \rVert_{\infty}^2 & ,
  \end{array}
}
where in the second line we  split the sum into contributions coming from  the different subspaces defined in eq.~\eqref{eq:Vheavylight}. 
In the third line we used $  \kappa_{-\lv}=( \kappa_{\lv})^{-1}$ and $ \kappa_{-\lv} \lVert  \hG_{-\ell} \rVert_{\infty}^2= \kappa_{\lv} \lVert  \hG_{\ell} \rVert_{\infty}^2$ (c.f.~eq.~\eqref{eq:ratiorhos}) to pair the contributions from $V_{\rm heavy}$ and $V_{\rm light}$ and make the summation explicitly in terms of elements of $V_{\rm heavy}$. As we saw in section \ref{nomodstabVrest}, fluxes in $V_{\rm rest}$ are either anti-self dual, or those in $V_0$ do not fix any moduli. Note that they do contribute to the tadpole, though. We thus get 
\eq{
  \label{tadpole_044d}
Q=\frac12  \langle G_4 , G_4 \rangle  \ge 
      \sum_{{\lv}>0} \gamma^{ \sum \lv_i}
  \lVert  \hG_{\ell} \rVert_{\infty}^2 \ .
  }
Since, again, there should at least be one $\hat G_{\ell}$ per moduli stabilized, this sum has at least $n_{\rm{mod}}$ terms, confirming the tadpole conjecture. Each term is weighted by a positive power of $\gamma$, which makes moduli stabilization in the asymptotic regime more difficult to achieve within the tadpole bound, even for a relatively small number of moduli  
(a similar observation was made in \cite{Betzler:2019kon}).
Note however that, as mentioned in the introduction, $\gamma$ need not be very large: $\gamma \gtrsim 4$ is enough for the sl(2) approximation to reproduce the actual vacua with high accuracy \cite{Grimm:2021ckh}. 


\subsubsection*{Quantization}

The norms $\lVert \hG_{\ell} \rVert^2_{\infty}$ appearing in \eqref{tadpole_044d} are in general 
not integer quantized and in principle could depend on the spectator moduli. It is therefore not obvious that the tadpole conjecture 
(reviewed in section~\ref{ss:tadpoleconjecture}) is satisfied
since  -- in an extreme situation -- the norms may scale as $\lVert \hG_{\ell} \rVert^2_{\infty}\sim 1/n_{\rm stab}$
and thereby violate the conjecture. However, we see no indications of such a scaling:
\begin{itemize}

\item 
In table~\ref{table:fluxquant} of appendix~\ref{app_flux_quant} we 
analyzed four examples with $n_{\rm stab}$ or order $20$. 
For each of these examples the spectator moduli decouple and the majority (more than $70\%$) of norms  of the subspaces satisfies 
$\lVert G_{\ell} \rVert^2_{\infty}\geq \tfrac{1}{4}$. Hence, there is no inverse scaling with 
the number of stabilized moduli $n_{\rm stab}$. 

\item Let us now clarify why our bounds on the norm of $G_\ell$, as opposed to the norm of the axion dependent $\hat{G}_\ell$ (which are the ones that appear in the definition of the tadpole, c.f. eq. \eqref{tadpole_044d}), are meaningful. First, note that the pairing that appears in the definition of the tadpole charge fulfills $\langle G_4, G_4\rangle = \langle \hat{G}_4, \hat{G}_4\rangle$, whereas this is not the case for the boundary norm, for which $\lVert G_4 \rVert_\infty\neq \lVert \hat{G}_4 \rVert_\infty$ in general (the equality holds if e.g.~all the axion vevs vanish). 
 
Nevertheless, 
the sum over $\ell$'s in equation \eqref{tadpole_044d} contains at least $n_{\rm stab}$ terms for which $\hat{G}_\ell=G_\ell$ (a single sl(2)-representation fixes a single modulus, see section \ref{sec:vacuawithVK3}) and the rest, if non-zero, give extra positive contributions.%
\footnote{Note also that, if an axion dependence is generated for a lower $\ell$ that happens to be populated also by the highest weight state in a different sl(2)-representation, there will be no mixing between the two because different sl(2)-representations are orthogonal.} Therefore, we can bound eq. \eqref{tadpole_044d}) as
\eq{
Q\geq \sum_{{\lv}>0} \gamma^{ \sum \lv_i}
  \lVert  \hG_{\ell} \rVert_{\infty}^2  \geq \sum_{{\lv}^{\prime}>0} \gamma^{ \sum \lv_i^{\prime}}
  \lVert  G_{\ell^{\prime}} \rVert_{\infty}^2 \ ,
  }
where the summation over $\ell^\prime$ indicates that only the fluxes in $\Vh$ that correspond to the highest-weight within each sl(2)-representation are included, for which $G_{\ell'}=\hG_{\ell'}$. 

\item The parameter $\gamma\gg 1$ parametrizes the strict asymptotic regime, and  in 
\cite{Grimm:2021ckh} it was found that $\gamma\gtrsim 4$ provides already a
good approximation of the moduli space structure for Calabi-Yau three-folds.
This parameter appears in \eqref{tadpole_044d} as $\gamma^{\sum \ell_i}$, where 
the sum is over positive integers taking values one or two. 
A very conservative estimate therefore is 
\eq{
  \gamma^{\sum \ell_i} \geq  \gamma \gtrsim 4 \,.
}
Furthermore, in our analysis in appendix~\ref{app_flux_quant}  we found that $70\%$ of the norms 
$\lVert G_{\ell} \rVert^2_{\infty}$ are larger than $1/4$. Using then that 
for stabilizing $n_{\rm stab}$ moduli there needs to be $n_{\rm stab}$ terms, 
we can  estimate
\eq{
  \sum_{{\lv}>0}   \lVert  \hG_{\ell} \rVert_{\infty}^2  > \frac{1}{4} \cdot 0.7 \cdot n_{\rm stab}\,.
}
However, our analysis 
provides only a lower bound on $\lVert G_{\ell} \rVert^2$ since we
determined the lowest non-zero values for each
$\lVert  G_{\ell} \rVert_{\infty}^2$ individually but did not account for 
the requirement that the corresponding fluxes have to stabilize
$n_{\rm stab}$ moduli.  (See the appendix 
for details on our analysis, in particular the discussion after \eqref{app_tadpole_001}). 
The values $\lVert G_{\ell} \rVert^2_{\infty}$ are therefore 
expected to be much larger.

\item Combining now the individual results above, we can 
give the following very conservative estimate for the tadpole charge 
\eq{
  Q > 0.7\op n_{\rm stab}\,,
  }
which is in agreement with the refined version of the Tadpole Conjecture, Eq. (\ref{refined}) (i.e. $\alpha>1/3$).

\end{itemize}


\section{Conclusions}
\label{sec:conclusions}

In this work we have studied the tadpole conjecture in asymptotic regions 
of complex structure moduli space by using the powerful tools of asymptotic 
Hodge theory. Our analysis was carried out in F-theory compactifications 
on Calabi-Yau four-folds in which the complex structure moduli are stabilized 
by a self-dual four-form flux $G_4$. We were able to give general 
evidence that the scaling 
of the tadpole with the number of stabilized directions is eminent if the moduli are 
stabilized in the strict 
asymptotic regions. These regions are close to the boundaries of the moduli space 
at which the Calabi-Yau space degenerates and one can additionally establish a hierarchy among the moduli values. 
Let us stress that apart from this hierarchy constraint our analysis was neither restricted to 
specific examples nor to specific asymptotic limits, such as the large complex structure 
limit. This generality puts much weight on the collected evidence and one can hope 
that the presented arguments are a first step to prove the conjecture in full generality. 

Our approach to the tadpole conjecture relied on the 
remarkable fact that in the strict asymptotic regimes the $(p,q)$-decomposition of the middle cohomology 
splits into representations of an sl$(2)^n$ algebra with commuting factors. 
Here the $n$ refers to the number of fields pushed to the asymptotic regime, which 
by assumption are the fields that we aim to stabilize in a vacuum. Switching on fluxes 
in the sl(2) eigenspaces we found that the commutativity of the sl(2)s reduces the generally 
complicated and coupled system of vacuum equations into a set of constraints that can 
be analyzed systematically. In particular, we showed that the Calabi-Yau condition ensures
that there is maximally a single sl$(2)$-representation with weights reaching from $-4$ to $4$. All 
other sl$(2)$-representations have a maximal weight $2$ and can formally be identified
with sl$(2)$-representations found for Calabi-Yau two-folds (K3 surfaces). For a large number $n$ of fields stabilized 
in the asymptotic regime we thus found that merely these K3-type representations are relevant
for stabilization. This has led to additional constraints that ensure, for example, that when stabilizing 
$n_{\rm stab}$ scalars the tadpole always admits order $n_{\rm stab}$ terms that grow in the vacuum expectation values of 
these fields. These findings give an explanation of the requirement in the tadpole conjecture 
to consider a large number of stabilized moduli. In fact, when considering only few moduli, the larger sl(2)-representations 
stemming from the existence of a (4,0)-form, and not being related to K3-representations, can be 
used to stabilize moduli and escape some of the stringent constraints. While at first counter-intuitive, 
we now see that as soon as we consider a large number of fields, we encounter more structure and constraints.
We believe that this feature persists when including corrections. 

Before turning to a brief discussion of the corrections to the strict sl(2)-splitting, let us stress an important 
subtle aspect that needs further clarification within our approach. It is known that the sl(2)-splitting of the flux cannot be generally performed over the 
integers but requires to use rational numbers. We have shown that in concrete examples the
denominators do not scale with the number of moduli, but note that an abstract analysis of the largest occurring denominator would require a more sophisticated mathematical argument which goes 
beyond the scope of this work. It would be desirable to address this question together with an in-detail analysis of moduli stabilization 
of axions $\phi^i$. The relation of these two issues arises from the fact that we have implemented the 
sl(2)-approximation also in the axion sector and worked with the axion-dependent flux $\hat G_4$. The 
latter links axion monodromies and fluxes and we expect additional constraints to arise from the integral 
quantization of fluxes. 
This highlights an important direction for future work.\footnote{In particular, let us stress 
that this quantization issue has appeared before in \cite{Gendler:2020dfp} in the analysis of the distance conjecture and weak gravity conjecture.}

The analysis of this paper concerns the moduli that are lifted in the strict asymptotic regime, for which we showed that whenever $n_{\text{stab}}$ moduli are stabilized by the leading contributions (i.e. ignoring the corrections), the tadpole grows linearly in the number of stablized moduli. An interesting challenge for future work is to extend our analysis to include corrections breaking the sl(2)-splitting and 
leave the strict asymptotic regime. In particular, one may want to keep the saxions $s^i$ large but allow for 
a stabilization without a hierarchy among the $s^i$. In this case polynomial corrections in the ratios of 
the $s^i$ will no longer be suppressed and play a major role in moduli stabilization. 
We expect that as one gets closer to the boundaries, the masses of moduli which are not stabilized in the strict asymptotic regime are asymptotically vanishing compared to those of moduli which are stabilized in the strict asymptotic regime.
For families of vacua close to the boundaries, the tadpole should then grow (at least) linearly with the number of moduli that remain stabilized at a hierarchically higher mass than the rest. Asymptotic 
Hodge theory provides powerful tools to systematically include such corrections (see, e.g.~\cite{Grimm:2020cda}). 
In the multi-moduli case, however, we expect that such an analysis will quickly get very involved. 
Nevertheless, such an extension will be essential to give a definite answer about the validity 
of the tadpole conjecture in asymptotic regimes. 
This extended analysis would then also cover the 
linear scenario presented in \cite{Palti:2008mg,Marchesano:2021gyv}, which might pose a challenge to this conjecture.\footnote{Note 
that it has recently been argued in \cite{Lust:2021xds,Grimm:2021ckh} (see also \cite{Plauschinn:2021hkp}) that there is no clear structural reason to expect a counter example
from the linear scenario.} An important feature of this stabilization scenario is that in the associated sl(2)-approximation 
a flat direction remains that then gets lifted after including corrections.

Let us also comment on the prospects of studying moduli stabilization and 
the tadpole conjecture in full generality. At first, one would expect that moduli 
stabilization is generic in this case, since even switching on a single flux quantum 
results in a highly non-trivial self-duality condition with polynomial and exponentially 
suppressed terms. This complexity is, however, deceiving when incorporating the 
results of the famous theorem about Hodge loci by Cattani, Deligne, and Kaplan \cite{CattaniDeligneKaplan}. 
Concretely, one can use this mathematical result to conclude that the locus in 
complex structure moduli space at which the flux $G_4$ is of type $(2,2)$ is actually given 
by algebraic equations. In other words, upon choosing appropriate coordinates, the moduli 
stabilization conditions are simply vanishing conditions on polynomials. While it is not known 
how the dimension of these spaces grows with the tadpole, it is conceivable from our 
asymptotic analysis that there is in fact a scaling with the number of moduli and it would be 
very interesting to prove such a scaling.  
Also allowing for a $(4,0)+(0,4)$ piece will, in general, 
destroy the algebraicity of the vacuum locus \cite{Bakker:2021uqw}. However, note that this generalization for a 
Calabi-Yau fourfold results in only a single complex equation independent of the number of 
moduli. This might indicate why in cases with few moduli the tadpole scaling can be violated 
while eventually it will be generally present when studying the stabilization of a 
large number of fields.


\vskip1em
\subsubsection*{Acknowledgments}
It is a pleasure to thank 
Brice Bastian,
Mehmet Demirtas, 
Arthur Hebecker,
\mbox{Stefano} Lanza, 
Chongchuo Li, 
Jeroen Monnee,
Andres Rios-Tascon, 
Lorenz \linebreak Schlechter, and 
Irene Valenzuela
for stimulating and very helpful discussions. 
TG and DH
are partly supported by the Dutch Research Council (NWO) via a Start-Up grant and a VICI grant, MG and AH are partly supported 
by the ERC Consolidator Grant 772408-Stringlandscape, and EP is
supported by a Heisenberg grant of the \textit{Deutsche Forschungsgemeinschaft} (DFG,
German Research Foundation) with project-number 430285316.


\clearpage
\appendix


\section{Asymptotic Hodge structures}
\label{app:AHS}
In this appendix we provide some background on the algebraic structures that underly strict asymptotic regimes: the sl(2)-decomposition and the boundary Hodge decomposition. In particular, we argue for the decomposition of the highest-weight states as described by \eqref{Eprimexp} given in the main text.


\subsubsection*{Pure Hodge structure}

Let us first recall the pure Hodge structure that lives in the strict asymptotic regime given by \eqref{Hodge sl2}. It is described by a Hodge decomposition of the primitive cohomology $H^4_{\rm prim}(Y_4)$ into $(p,q)$-form pieces as
\begin{equation}
H^4_{\rm prim}(Y_4) = \bigoplus_{p+q=4} H^{p,q}_{\rm sl(2)}\, ,
\end{equation}
where $\bar{H}^{p,q}_{\rm sl(2)} = H^{q,p}_{\rm sl(2)}$. In order to make the underlying boundary structures more precise, it is helpful to recast this splitting in terms of a so-called Hodge filtration $F^p$. These vector spaces collect the $(p,q)$-eigenspaces of the Hodge decomposition as
\begin{equation}
F^p_{\rm sl(2)} = \sum_{q \geq p}H^{q,4-q}_{\rm sl(2)}\, .
\end{equation}
This filtration  varies holomorphically in the complex structure moduli $t^i$. By the sl(2)-approximation it can be described as
\begin{equation}
F^p_{\rm sl(2)}(t) =e^{t^i N_i^-} F^p_{(n)}\, ,
\end{equation}
where the limiting filtration $F^p_{(n)}$ only depends on moduli not taken to the boundary. For later reference, let us note that we can generate $n$ other limiting filtrations $F^p_{(k)}$ through the recursion relation
\begin{equation}
F^p_{(k)} = e^{i N^-_{k+1}} F^p_{(k+1)} .
\end{equation}


\subsubsection*{Mixed Hodge structure}

Let us point out that the filtration $F^p_{(0)}$ obtained in this way is precisely the Hodge decomposition associated to the boundary described by \eqref{eq:Hpqinfty}. For the other limiting filtrations there generically is no notion of a pure Hodge structure; instead these give rise to mixed Hodge structures, which can be made precise through Deligne splittings. These first require us to introduce monodromy weight filtrations $W_{\ell}$, which are vector spaces constructed out of the kernels and images of the lowering operators as
\begin{equation}
W_\ell(N) =\!\!\! \sum_{j \geq \max(-1, \ell-4)} \ker N^{j+1} \cap \text{img} \op N^{j-\ell+4}\, , \qquad N=N^-_{(k)} = N_1^- + \ldots N_k^-\, , 
\end{equation}
with $N^-_{(0)}=0$. The Deligne splitting describing the mixed Hodge structure at step $k$ is then given by
\begin{equation}
I^{p,q}_{(k)} = F^p_{(k)} \cap \overline{F}^q_{(k)} \cap W_{p+q}(N^-_{(k)})\, .
\end{equation}
We stress that we already specialized to the sl(2)-splitting from the beginning in this paper, and in general other pieces have to be taken into account in this intersection of vector spaces altering the conjugation property $\overline{I}^{p,q}=I^{q,p}$. The weight operators of the sl(2)-triples are then understood as multiplying an element by its row $p+q$ as
 \begin{equation}
 v_{p,q} \in I^{p,q}_{(k)}: \quad N^0_{(k)} v_{p,q} = (p+q-4) v_{p,q}\, .
 \end{equation} 
Of special importance to us are the highest-weight components under the lowering operators $N^-_{(k)}$, which can be defined as
\begin{equation}
P^{p,q}_{(k)} = I^{p,q}_{(k)} \cap \ker[ (N^-_{(k)})^{p+q-3} ]\, .
\end{equation}
The Deligne splitting can then be recovered from the highest-weight subspaces and their descendants as
\begin{equation}
I^{p,q}_{(k)} = \bigoplus_{r} (N^-_{(k)})^k P^{p+r, q+r}_{(k)}\, .
\end{equation}


\subsubsection*{Allowed weights}

Having introduced the Deligne splittings and their highest-weight decompositions, we are finally in the position to look more closely at the allowed weights for highest-weight states, i.e.~the splitting into the $(4,0)$-form part and K3 Hodge structures given in \eqref{H4decompK3}. To this end, the crucial relation between the highest-weight subspaces is given by
\begin{equation}
e^{i N^-_{k+1}} P^{p,q}_{(k+1)} \subseteq P^{p,q-\ell_{k+1}+\ell_k}_{(k)} \, ,
\end{equation}
assuming $p \geq q$. Pictorially this relation can be interpreted as follows: a highest-weight state starting at position $(p,q)$ in the Deligne diamond can only end up in another position $(p,q')$ with $q' \geq q$ and $q' \leq p$. In other words, only diagonal displacements to right-above are allowed, and never across the middle. This rule is rather abstract, so let us draw it explicitly for the Deligne diamond of a Calabi-Yau fourfold
  \begin{equation}\label{diamond}
  \begin{array}[c]{c}
 \begin{tikzpicture}[scale=1,cm={cos(45),sin(45),-sin(45),cos(45),(15,0)}]
  \draw[step = 1, gray, ultra thin] (0, 0) grid (4, 4);
  
\draw[fill, blue] (0,4) circle[radius=0.05] node[left]{\scriptsize $1$};
\draw[fill, blue] (1,4) circle[radius=0.05] node[left]{\scriptsize  $d_i = 1$};
\draw[fill, blue] (2,4) circle[radius=0.05] node[left]{\scriptsize $d_i = 2$};
\draw[fill, blue] (3,4) circle[radius=0.05] node[left]{\scriptsize $d_i = 3$};
\draw[fill, blue] (4,4) circle[radius=0.05] node[left]{\scriptsize $d_i = 4$};

\draw[fill, red] (1,3) circle[radius=0.05] node[left]{\scriptsize $h^{3,1}$};
\draw[fill, red] (2,3) circle[radius=0.05] node[left]{\scriptsize $\ell_i = 1$};
\draw[fill, red] (3,3) circle[radius=0.05] node[left]{\scriptsize $\ell_i = 2$};

\draw[fill, green] (2,2) circle[radius=0.05] node[left]{\scriptsize $h^{2,2}$};

\draw[->, blue] (0.1,4) -- (0.9 ,4);
\draw[->, blue] (1.1,4) -- (1.9 ,4);
\draw[->, blue] (2.1,4) -- (2.9 ,4);
\draw[->, blue] (3.1,4) -- (3.9 ,4);

\draw[->, red] (1.1,3) -- (1.9 ,3);
\draw[->, red] (2.1,3) -- (2.9 ,3);
\end{tikzpicture} 
\end{array}
 \end{equation}
We indicated the (up to rescaling) unique state starting in $I^{4,0}_{(0)}$ -- corresponding to the leading term $\tilde a_0$ of the $(4,0)$-form periods -- in blue, while the remainder of the diagram has been highlighted in red and green. Note also that, in principle, a highest weight state is allowed to skip in-between steps, e.g.~it can move directly from $d_1=1$ to $d_2=4$. From this decomposition we see that: (1) there is one state corresponding to the outer blue part of the diamond, (2) there are $h^{3,1}$ states that can move up at most two steps in red, and (3) the remainder of the highest-weight states in green is fixed at $I^{2,2}$. This matches precisely with the allowed indices given in \eqref{Eprimexp}. 


\section{Bases of sl(2)-representations and  $\star_\infty$ action}
\label{app:Cinfty}

We include here some useful explicit expressions and properties the basis vectors for the subspaces $\Vell$, including the explicit action of $\star_{\infty}$ and the lowering operators. Recall that the pairing $\langle \cdot \,  ,  \, \cdot \rangle$ fulfills
\begin{equation}
 \langle \;\cdot\; , N_i^- \;\cdot\; \rangle=-\langle  N_i^-  \; \cdot  \; , \; \cdot \;\rangle \, .
 \end{equation}


\subsubsection*{The space $P_{0}$}

The space $P_0$ is formed by $P_0^{3,1}$, $P_0^{1,3}$, and $P_0^{2,2}$, according to their decomposition in the last Hodge-Deligne diamond. All of them are annihilated by all the $N_i^-$. The former two form a (real) two-dimensional space, $P_0^{3,1} \oplus P_0^{1,3}$. A real basis can be obtained from the real and imaginary parts of the complex basis vector in $P_0^{3,1}$, which we denote
\begin{equation}
v_0^R \, , \qquad v_0^I,
\end{equation}
with the non-vanishing pairings
\begin{equation}
\langle v_0^R , v_0^R \rangle = -1\, , \qquad \langle v_0^I , v_0^I \rangle = -1\, .
\end{equation}
The action of $\star_\infty$ is 
\begin{equation}
\star_\infty v_0^R=- v_0^R \, , \qquad \star_\infty v_0^I=- v_0^I\, .
\end{equation}

The space $P_0^{2,2}$ is one dimensional, with the pairing and the action of the boundary Hodge star on the basis vector given by
\begin{equation}\label{eq:pairingCinftyP022}
\langle v_0^{(2,2)} , v_0^{(2,2)} \rangle = 1 \, , \qquad \star_\infty v_0^{(2,2)}=v_0^{(2,2)}\, .
\end{equation} 


\subsubsection*{The spaces generated by $P_{01_i}$}
Each of these spaces is a complex space of dimension two, since it includes both the $P_{01_i}^{3,2}$ and $P_{01_i}^{2,3}$ in the last Hodge-Deligne diagram associated to the real splitting. It consists of $V_{01_i}$ and $V_{0-1_i}$, each of them of real dimension two. We can define a real basis by taking the real and imaginary parts of the complex vector that sits in $P_{01_i}^{3,2}$ and applying $N_i^-$ to each of them separately.
\begin{equation}
\label{eq:basisP01}
v_{01_i}^{R} \, , \qquad v_{01_i}^{I} \, , \qquad  v_{0-1_i}^{R}=N_i^- v_{01_i}^{R} \, \qquad v_{0-1_i}^{I}=N_i^- v_{0-1_i}^{I}   \, ,
\end{equation}
with the following non-vanishing pairings
\begin{equation}
\langle v_{01_i}^{R} ,  v_{0-1_i}^{I} \rangle = 1 \,,  \hspace{50pt} \langle v_{01_i}^{I} ,  v_{0-1_i}^{R} \rangle = -1 \, .
\end{equation} 
The action of $\star_{\infty}$ on the basis vectors is
\begin{equation}
\begin{array}{lcl}
\star_{\infty} v_{01_i}^{R}=+v_{0-1_i}^{I}\, , && 
\star_{\infty} v_{0-1_i}^{I}=+v_{01_i}^{R}\, , \\[4pt]
\star_{\infty} v_{01_i}^{I}=-v_{0-1_i}^{R}\,  , &&
\star_{\infty} v_{0-1_i}^{R}=-v_{01_i}^{I}\, .
\end{array}
\end{equation}


\subsubsection*{The spaces generated by $P_{02_i}$}
Each of these spaces has (real) dimension three, and it consists on $V_{02_i} \, \oplus \,  N^-_i P_{02_i} \, \oplus \, V_{0-2_i}$ as defined in eq.~\eqref{eq:Vellspaces}, where we recall that $ N^-_i P_{02_i}$ includes the part of $V_0$ that is not highest-weight. We introduce the following basis vector for each of these $\Vell$, respectively
\begin{equation}
\label{eq:basisP02}
v_{02_i}\, ,  \qquad v_{0}= N_i^- v_{02_i} \, , \qquad v_{0-2_i}= (N_i^-)^2 v_{02_i}\, .
\end{equation}
Their non-zero inner products are 
\begin{equation}
\langle v_{02_i}, v_{0-2_i} \rangle=1\,, \hspace{50pt} \langle v_{0}, v_{0}\rangle = -1 \, ,
\end{equation} 
and the action of $\star_\infty$ on each of them takes the form
\begin{equation} \label{eq:CinftyP02}
\star_{\infty} v_{02_i}=\dfrac{1}{2}v_{0-2_i} \,, \qquad
\star_{\infty}v_{0}=-v_0 \,, \qquad 
\star_{\infty}v_{0-2_i}=2 \, v_{02_i}\, .
\end{equation}


\subsubsection*{The spaces generated by $P_{01_i2_j}$}
In this case, each of these subspaces has (real) dimension four, and it consists of $V_{01_i2_j} \, \oplus \, V_{01_i0_j} \, \oplus \, V_{0-1_i0_j}    \, \oplus \, V_{0-1_i-2j}$ (c.f.~ eq.~\eqref{eq:Vellspaces}). Introducing the  basis vectors
\begin{equation}
\label{eq:basisP012}
v_{01_i2_j} \, , \hspace{15pt} 
v_{01_i0_j}=N_j^- v_{01_i2_j}\, ,  \hspace{15pt}
v_{0-1_i0_j}=N_i^- v_{01_i2_j} \, ,  \hspace{15pt} 
v_{0-1_i-2j}=N_i^- \, N_j^- v_{01_i2_j} \, ,
\end{equation}
we obtain the following non-vanishing pairings
\begin{equation}
\langle v_{01_i2_j} ,v_{0-1_i-2_j} \rangle = 1 \,, \hspace{50pt} \langle v_{01_i0_j} ,v_{0-1_i0_j} \rangle =- 1 \, .
\end{equation}
Finally, the $\star_{\infty}$ operator acts as
\begin{equation}
\label{eq:CinftyP012}
\begin{array}{l c l}
\star_{\infty} v_{01_i2_j}=+v_{0-1_i-2_j}\, , & &
\star_{\infty} v_{0-1_i-2_j}=+ v_{01_i2_j}\,  ,\\[4pt]
\star_{\infty} v_{01_i0_j}=-v_{0-1_i0_j}\, , &  &
\star_{\infty} v_{0-1_i0_j}=-v_{01_i0_j}\, .
\end{array}
\end{equation}


\section{Weak coupling-conifold example}\label{ss:2moduliexample}
\label{app:example}

In this appendix we consider a simple two-moduli example, where one complex structure modulus is sent to weak coupling and the other towards a conifold point. The first corresponds to Sen's limit \cite{Sen:1996vd} and this modulus can be understood as the axio-dilaton of Type IIB; the second is the conifold modulus of the Calabi-Yau threefold in this setup. This example highlights that asymptotic Hodge theory can be applied near any boundary in complex structure moduli space, allowing us to probe regions away from large complex structure. More concretely, it illustrates how the sl(2)-decomposition splits into representations associated to the (4,0)-form and a K3 subblock. This example demonstrates the interplay between fluxes in different representations in the stabilization of complex structure moduli.

\subsubsection*{Period vector data}
We begin by describing the period vector of the (4,0)-form near such a weak coupling-conifold point. Let us point out that we do not refer to any explicit geometrical example here, but merely use the typical form of the periods in such a regime. The leading part of the period vector takes the form
\begin{equation}
\begin{aligned}
\Pi &=\Big(
1+ \frac{a^2 z^2}{8 \pi }, \ a z, \  i-\frac{i a^2 z^2}{8 \pi }, \ \frac{i a z (\log (z)-1)}{2 \pi } , \\
&\hspace{40pt} \tau  \big(1+\frac{a^2 z^2}{8 \pi }\big), \ a \tau  z, \ \tau  \big(i-\frac{i a^2 z^2}{8 \pi }\big), \ \frac{i a \tau  z (\log (z)-1)}{2 \pi } \Big)\,,
\end{aligned}
\end{equation}
where $0 \neq a \in \mathbb{R}$, with the conventions for the conifold periods used in \cite{Bastian:2021eom}. The axio-dilaton is denoted by $\tau=c+i s$, and the conifold modulus by $z = e^{2\pi i t} = e^{2\pi i (x+i y)}$. The period vector can be brought to the standard form of the nilpotent orbit approximation as\footnote{The terms proportional to $a_{01}$ and $a_{02}$ are the exponential corrections to the leading order term, that we did not display in the main text (see equation \ref{eq:pisl2}). These are essential for computing a non-degenerate Hodge star.} 
\begin{equation}
\Pi = e^{\tau  N_1+t N_2} (a_{00}+e^{2\pi i t} a_{01} + e^{4\pi i t} a_{02})\, , 
\end{equation}
where the log-monodromy matrices $N_i$ are given by
\begin{equation}\label{eq:exampleNs}
N_1 = \scalebox{0.8}{$\left(
\begin{array}{cccccccc}
 0 & 0 & 0 & 0 & 0 & 0 & 0 & 0 \\
 0 & 0 & 0 & 0 & 0 & 0 & 0 & 0 \\
 0 & 0 & 0 & 0 & 0 & 0 & 0 & 0 \\
 0 & 0 & 0 & 0 & 0 & 0 & 0 & 0 \\
 1 & 0 & 0 & 0 & 0 & 0 & 0 & 0 \\
 0 & 1 & 0 & 0 & 0 & 0 & 0 & 0 \\
 0 & 0 & 1 & 0 & 0 & 0 & 0 & 0 \\
 0 & 0 & 0 & 1 & 0 & 0 & 0 & 0 \\
\end{array}
\right)$}\, , \quad  N_2 = \scalebox{0.8}{$\left(
\begin{array}{cccccccc}
 0 & 0 & 0 & 0 & 0 & 0 & 0 & 0 \\
 0 & 0 & 0 & 0 & 0 & 0 & 0 & 0 \\
 0 & 0 & 0 & 0 & 0 & 0 & 0 & 0 \\
 0 & -1 & 0 & 0 & 0 & 0 & 0 & 0 \\
 0 & 0 & 0 & 0 & 0 & 0 & 0 & 0 \\
 0 & 0 & 0 & 0 & 0 & 0 & 0 & 0 \\
 0 & 0 & 0 & 0 & 0 & 0 & 0 & 0 \\
 0 & 0 & 0 & 0 & 0 & -1 & 0 & 0 \\
\end{array}
\right)$}\, ,
\end{equation}
and the terms $a_{ij}$ in the expansion of the period vector read 
\begin{equation}
\begin{aligned}
a_{00} &= \big(1, \ 0, \ i, \ 0, \ 0, \ 0, \ 0, \ 0 \big)\, , \quad a_{01} = a \big(0, \ 1, \ 0, \ -\frac{i}{2\pi }, \ 0, \ 0, \ 0, \ 0 \big)\, , \\
 a_{02} &=  \frac{a^2}{4\pi}\big(1, \ 0, \ -i, \ 0, \ 0, \ 0, \ 0, \ 0 \big)\, .
 \end{aligned}
\end{equation}
For later reference, let us write down the expression for the bilinear pairing  defined in \eqref{norm&ip}
\begin{equation}
\eta = \scalebox{0.8}{$\left(
\begin{array}{cccccccc}
 0 & 0 & 0 & 0 & 0 & 0 & -1 & 0 \\
 0 & 0 & 0 & 0 & 0 & 0 & 0 & -1 \\
 0 & 0 & 0 & 0 & 1 & 0 & 0 & 0 \\
 0 & 0 & 0 & 0 & 0 & 1 & 0 & 0 \\
 0 & 0 & 1 & 0 & 0 & 0 & 0 & 0 \\
 0 & 0 & 0 & 1 & 0 & 0 & 0 & 0 \\
 -1 & 0 & 0 & 0 & 0 & 0 & 0 & 0 \\
 0 & -1 & 0 & 0 & 0 & 0 & 0 & 0 \\
\end{array}
\right)$}\, .
\end{equation}
Given this data one has to compute the sl(2) approximation (see details in \cite{Grimm:2021ckh,Grimm:2019ixq}). The lowering operators $N_1^-=N_1$ and $N_2^-=N_2$ in \eqref{eq:exampleNs} are completed into sl(2)-triples by weight operators
\begin{equation}
N_1^0 =  \scalebox{0.8}{$\left(
\begin{array}{cccccccc}
 1 & 0 & 0 & 0 & 0 & 0 & 0 & 0 \\
 0 & 1 & 0 & 0 & 0 & 0 & 0 & 0 \\
 0 & 0 & 1 & 0 & 0 & 0 & 0 & 0 \\
 0 & 0 & 0 & 1 & 0 & 0 & 0 & 0 \\
 0 & 0 & 0 & 0 & -1 & 0 & 0 & 0 \\
 0 & 0 & 0 & 0 & 0 & -1 & 0 & 0 \\
 0 & 0 & 0 & 0 & 0 & 0 & -1 & 0 \\
 0 & 0 & 0 & 0 & 0 & 0 & 0 & -1 \\
\end{array}
\right)$}\, , \quad N_2^0 =  \scalebox{0.8}{$\left(
\begin{array}{cccccccc}
 0 & 0 & 0 & 0 & 0 & 0 & 0 & 0 \\
 0 & 1 & 0 & 0 & 0 & 0 & 0 & 0 \\
 0 & 0 & 0 & 0 & 0 & 0 & 0 & 0 \\
 0 & 0 & 0 & -1 & 0 & 0 & 0 & 0 \\
 0 & 0 & 0 & 0 & 0 & 0 & 0 & 0 \\
 0 & 0 & 0 & 0 & 0 & 1 & 0 & 0 \\
 0 & 0 & 0 & 0 & 0 & 0 & 0 & 0 \\
 0 & 0 & 0 & 0 & 0 & 0 & 0 & -1 \\
\end{array}
\right)$}\, , 
\end{equation}
which indeed fulfill the standard commutation relations \eqref{aht_001}. For our purposes we will not need the raising operators, but in principle one could obtain these by solving the remaining commutation relations.

\subsubsection*{Highest-weight states and Hodge decompositions}
Having set up the period vector and sl(2)-algebras, we next study the decomposition of the middle cohomology $H^4_{\rm prim}(Y_4)$ in the strict asymptotic regime under the sl(2)-approximation. We begin with the highest-weight states \eqref{eq:primitive} of the sl(2)-representations. Two highest-weight states are obtained from $a_0$ and its conjugate, which have weights $\ell=(1,1)$ under $N_{1}^0$ and $N_1^0+N_2^0$. The remaining highest-weight state is given by $a_{01}+\frac{i}{2\pi } N_2 a_{01}$, which has weights $\ell=(1,2)$. Altogether we find
\begin{equation}
\begin{aligned}
P_{11}^{4,1}&: \quad \big(1, \ 0, \ i, \ 0, \ 0, \ 0, \ 0, \ 0 \big)\, , \\
P_{12}^{3,3}&: \quad  \big(0, \ 1, \ 0, \ 0, \ 0, \ 0, \ 0, \ 0 \big)\, , \\
P_{11}^{1,4}&: \quad \big(1, \ 0, \ -i, \ 0, \ 0, \ 0, \ 0, \ 0 \big)\, . \\
\end{aligned}
\end{equation}
These sl(2)-representations are completed by considering descendants under the action of $N_1,N_2$, which are given by
\begin{equation}
\begin{aligned}
N_1 P_{11}^{4,1} \subset V_{-1-1} &: \quad \big(0, \ 0, \ 0, \ 0, \ 1, \ 0, \ +i, \ 0 \big)\,, \\
N_1 P_{11}^{1,4} \subset V_{-1-1} &: \quad \big(0, \ 0, \ 0, \ 0, \ 1, \ 0, \ -i, \ 0 \big)\,, \\
N_1 P_{12}^{3,3} \subset V_{-10} &: \quad \big(0, \ 0, \ 0, \ 0, \ 0, \ 1, \ 0, \ 0 \big)\,, \\
N_2 P_{12}^{3,3} \subset V_{10} &:\quad \big(0, \ 0, \ 0, \ 1, \ 0, \ 0, \ 0, \ 0 \big)\, , \\
N_1 N_2 P_{12}^{3,3} \subset V_{-1-2} &:\quad \big(0, \ 0, \ 0, \ 0, \ 0, \ 0, \ 0, \ 1 \big)\, . \\
\end{aligned}
\end{equation}
Recalling the splitting \eqref{H4decompK3} of the sl(2)-representations we then decompose the middle cohomology into two parts as 
\begin{equation}
H^{4}_{\rm prim}(Y_4, \mathbb{C}) = H_\Omega \oplus H_{{\rm K3}}\, ,
\end{equation}
where we defined the terms
\begin{equation}
\begin{aligned}
H_\Omega &= P_{11}^{4,1} \oplus N_1 P_{11}^{4,1} \oplus P_{11}^{1,4} \oplus N_1 P_{11}^{1,4} \, , \\
H_{{\rm K3}} &= P_{12}^{3,3} \oplus N_1 P_{12}^{3,3}  \oplus N_2 P_{12}^{3,3}  \oplus N_1 N_2 P_{12}^{3,3}\, .
\end{aligned}
\end{equation}
Note that $H_\Omega$ is spanned by vectors with non-vanishing entries in odd positions, while $H_{{\rm K3}}$ is spanned by vectors with non-vanishing entries in even positions. 

We next consider the Hodge decomposition of these individual terms $H_\Omega$ and $H_{{\rm K3}}$ in the strict asymptotic regime. As Hodge structure on the $(4,0)$-form representations we find
\begin{equation}
H_\Omega = H^{4,0}_\Omega \oplus H^{3,1}_\Omega \oplus H^{1,3}_\Omega \oplus H^{0,4}_\Omega\, ,
\end{equation}
with subspaces spanned by 
\begin{equation}\label{eq:OmegaHodge}
\begin{aligned}
H^{4,0}_\Omega &: \quad \big( 1, \ 0,\ i,\ 0,\ c+is , \ 0,\  i (c+is) ,\  0 \big)\, , \\
H^{3,1}_\Omega &: \quad \big(1, \ 0,\  i,\  0,\ c - i s,\ 0,\ i (c -i s),\ 0 \big)\, .
\end{aligned}
\end{equation}
and the others determined by complex conjugation. The $(4,0)$-form subspace is straightforwardly determined as $H^{4,0}_\Omega = e^{t^i N_i} P_{11}^{4,1}$ from the highest-weight subspace according to \eqref{eq:pqformsl2}; the $(3,1)$-form subspace is spanned by the linear combination of $\Pi_{\rm sl(2)}$ and $\partial_\tau \Pi_{\rm sl(2)}$ that is orthogonal to $\overline{\Pi}_{\rm sl(2)}$ under the bilinear pairing (which is precisely the one picked by the K\"ahler covariant derivative).
The other part of the cohomology corresponds to the K3 block: its Hodge decomposition takes the form
\begin{equation}
H_{{\rm K3}} = (H_{{\rm K3}} )^{3,1} \oplus (H_{{\rm K3}} )^{2,2} \oplus (H_{{\rm K3}} )^{1,3}\, ,
\end{equation}
with subspaces spanned by
\begin{equation}\label{eq:K3Hodge}
\begin{aligned}
(H_{{\rm K3}} )^{3,1} &: \quad \big( 0, \ 1, \ 0, \ -(x+iy), \ 0, \ c+is , \ 0, \ -(x+iy)(c+is) \big)\, , \\
(H_{{\rm K3}} )^{2,2} &: \quad \big( 0, \ y, \ 0, \ 0, \ 0, \ cy+sx, \ 0, \ - s (x^2+y^2) \big)\, , \\
& \quad \ \ \ \big( 0, \ 0, \ 0, \ y, \ 0, \ s, \ 0, \ cy-sx\big)\, , \\
\end{aligned}
\end{equation}
and the $(1,3)$-form subspace fixed by complex conjugation. Now the $(3,1)$-form subspace is straightforwardly determined as $(H_{{\rm K3}} )^{3,1}  = e^{t^i N_i} P_{12}^{3,3}$ from the highest-weight subspace according to \eqref{eq:pqformsl2}; the $(2,2)$-subspace is determined as the part of $H_{{\rm K3}}$ which is orthogonal to the $(3,1)$- and $(1,3)$-subspaces under the bilinear pairing. 
Note that in this example where the number of moduli is not large, the subspaces $H_{\Omega}$ and $H_{\rm K3}$ have roughly the same dimensions, while for large moduli the dimension of $H_{\Omega}$ is ${\cal O}(1)$ while the dimension of $H_{\rm K3}$ is ${\cal O}(n)$.

Putting the above two Hodge decompositions \eqref{eq:OmegaHodge} and \eqref{eq:K3Hodge} together, we can write the sl(2)-approximated Hodge star operator of the middle cohomology $H^4_{\rm prim}(Y_4, \mathbb{C})$ as
\begin{equation}
\star_{\rm sl(2)} = \scalebox{0.8}{$\left(
\begin{array}{cccccccc}
 0 & 0 & \frac{c}{s} & 0 & 0 & 0 & -\frac{1}{s} & 0 \\
 0 & \frac{c x}{s y} & 0 & \frac{c}{s y} & 0 & -\frac{x}{s y} & 0 &
   -\frac{1}{s y} \\
 -\frac{c}{s} & 0 & 0 & 0 & \frac{1}{s} & 0 & 0 & 0 \\
 0 & -\frac{c \left(x^2+y^2\right)}{s y} & 0 & -\frac{c x}{s y} & 0 &
   \frac{x^2+y^2}{s y} & 0 & \frac{x}{s y} \\
 0 & 0 & \frac{c^2}{s}+s & 0 & 0 & 0 & -\frac{c}{s} & 0 \\
 0 & \frac{x \left(c^2+s^2\right)}{s y} & 0 & \frac{c^2+s^2}{s y} & 0 &
   -\frac{c x}{s y} & 0 & -\frac{c}{s y} \\
 -\frac{c^2+s^2}{s} & 0 & 0 & 0 & \frac{c}{s} & 0 & 0 & 0 \\
 0 & -\frac{\left(c^2+s^2\right) \left(x^2+y^2\right)}{s y} & 0 & -\frac{x
   \left(c^2+s^2\right)}{s y} & 0 & \frac{c \left(x^2+y^2\right)}{s y} & 0
   & \frac{c x}{s y} \\
\end{array}
\right)$}\, .
\end{equation}
One can verify straightforwardly that $\star_{\rm sl(2)}$ acts as $(-1)^{(p-q)/2}$ on elements of the subspaces $(H_{{\rm K3}} )^{p,q}$ and $H_\Omega^{p,q}$.

\subsubsection*{Self-duality conditions and flux vacua}
Having constructed the sl(2)-approximation and corresponding Hodge star operator, we next study the self-duality condition for the fluxes in the strict asymptotic regime. We treat $H_\Omega$ of the $(4,0)$-form and the K3 block $H_{{\rm K3}}$ individually. Let us denote the four-form flux by a vector as $G_4 = (h_1,h_2,h_3,h_4,f_1,f_2,f_3,f_4)$. The self-duality condition on the subspace $H_\Omega$ then reads
\begin{equation}
-f_3+ch_3-h_1s=0\, , \qquad f_1-ch_1-h_3 s = 0\, .
\end{equation}
Note that these constraints can equivalently be obtained by demanding orthogonality under the bilinear pairing with the $(3,1)$-form subspace given in \eqref{eq:OmegaHodge}, since the $(4,0)$-form subspace and its conjugate are self-dual. It is solved by
\begin{equation}\label{eq:exampletau}
c = \frac{f_1 h_1+f_3 h_3}{h_1^2+h_3^2}\, , \qquad s = \frac{-f_3 h_1+f_1 h_3}{h_1^2+h_3^2}\, .
\end{equation}
We next turn to the self-duality condition of the K3 subblock $H_{{\rm K3}}$. Recall from \eqref{eq:H31ortho} that it is most conveniently imposed by demanding orthogonality with the $(3,1)$-form subspace. Taking the $(3,1)$-form given in \eqref{eq:K3Hodge} we obtain as condition
\begin{equation}\label{eq:example31}
(h_2 \tau -f_2)(h_2 t +h_4) = h_2 f_4 - f_2 h_4\, .
\end{equation}
The flux quanta here are identified with those in \eqref{eq:G4V012} as
\begin{equation}
G_{12}=h_2, \quad G_{-10} = f_2, \quad G_{10} =- h_4, \quad G_{-1-2}=-f_4\, .
\end{equation}
The solution to \eqref{eq:example31} is described by some simple complex function $t(\tau)$. In \eqref{eq:minP012_2simple} this function parametrized the flat direction of the scalar potential; here these are lifted by the inclusion of fluxes in $H_\Omega$, which fixes $\tau$ by \eqref{eq:exampletau}, and thus also $t$.


\section{Tadpole contribution and flux quantization}
\label{app_flux_quant}

In this appendix we investigate how the quantization of fluxes affects the tadpole for a few examples. To be precise, we consider Calabi-Yau three-fold geometries near the large complex structure regime with large $h^{2,1}$.\footnote{For simplicity we use Calabi-Yau three-folds but we do not expect the quantization issues to be very different in four-folds, in particular the absence of inverse scaling with the number of moduli, as we will show.} By using mirror symmetry we compute the relevant data with CYTools \cite{cytools}.


\subsubsection*{Quantization of sl(2)-eigenspaces}

Let us first recall the form of the tadpole in strict asymptotic regimes as given in \eqref{tadpole_044d}. We want to identify the relevant quantity to check for flux quantization, and make sure that there are no pieces that can scale inversely with the number of moduli. We decomposed the tadpole contribution of the fluxes as
\begin{equation}\label{eq:apptadpole}
\langle G_4 , G_4 \rangle \geq 2 \sum_{\ell >0} \gamma^{\sum \ell_i} \| \hat{G}_\ell \|_\infty^2\, .
\end{equation}
This sum runs over at least as many terms as the number of stabilized moduli $n_{\rm stab}$. Therefore, in order for the tadpole to grow linearly with $n_{\rm stab}$, we have to require that each of these terms is order one. The strict asymptotic regime already requires $\gamma \gg 1$, so the problem at hand reduces to checking whether the boundary norm $\| \hat{G}_\ell \|_\infty^2$ defined in \eqref{eq:boundarynorm} is of order one. We explicitly derive lower bounds for these coefficients for each of the heavy sl(2)-eigenspaces $V_{\ell}$ in a few three-fold examples, given in table \ref{table:fluxquant}.

\begin{table}[t]
\centering
\renewcommand*{\arraystretch}{1.8}
\begin{tabular}{|c || c | c | c | c | c |}
\hline & $h^{1,1}$ & $n_{\rm gen}$ & $n_{\rm stab}$ & heavy $V_\ell$ & $\| {G}_\ell \|_\infty^2 \geq \frac{1}{4}$ \\ \hline \hline 
example 1 & 70 & 53 & 20 & 21 & 15 \\ \hline
example 2 & 75 & 65 & 20 & 21 & 19 \\ \hline
example 3 & 100 & 47 & 24 & 25 & 21\\ \hline
example 4 & 100 & 67 & 22 & 23 & 18\\ \hline
\end{tabular}
\caption{\label{table:fluxquant}In this table we summarize the statistics of our study of geometries with large $h^{1,1}$: the number $n_{\rm gen}$ indicates the number of generators we managed to compute for the K\"ahler cone; the number $n_{\rm stab}$ gives the number of moduli in which the sl(2)-approximated Hodge star $\star_{\rm sl(2)}$ varies in the strict asymptotic regime we chose; the next column specifies the number of sl(2)-eigenspaces $V_{\ell}$ which are heavy asymptotically; the last column indicates how many out of these heavy $V_{\ell}$ have a boundary Hodge norm bounded from below by at least $1/4$.
}
\end{table}


\subsubsection*{Details of the computation}

We now provide some details of how the calculation above was implemented. 

\begin{itemize}

\item We compute the relevant topological data of the examples using CYTools. We take four examples at various large values of $h^{1,1}$ in the Kreuzer-Skarke dataset \cite{Kreuzer:2000xy}. The intersection numbers of a given geometry are efficiently computed using this program, as well as the generators of the Mori cone.\footnote{To be more precise, let us note that it is much easier in practice to compute the Mori and K\"ahler cone for the ambient space rather than the Calabi-Yau manifold itself. The K\"ahler cone of the ambient space is in general contained in the K\"ahler cone of the Calabi-Yau manifold, so we thereby restrict our attention to a smaller portion of the moduli space. We refer to \cite{Altman:2014bfa,Demirtas:2018akl, Demirtas:2020dbm} for more details on constructions of Calabi-Yau hypersurfaces in toric ambient spaces.} However, for our purposes we need to know generators $\omega_i$ of the K\"ahler cone: in this basis we can expand the K\"ahler form as $J = t^i \omega_i$, where sending the saxion $\text{Im} \, t^i = s^i$ to infinity corresponds to the large field limit to the boundary. The computational cost of dualizing a generic Mori cone to the K\"ahler cone scales exponentially with $h^{1,1}$, and  for $h^{1,1} \gtrsim 20$ it becomes essentially impossible to compute all of these generators. 

\item Let us first briefly review some key aspects of dualizing a Mori cone $\cM$ to the K\"ahler cone $\cK$, and in particular highlight the complications that arise at large $h^{1,1}$. We write the generators of the Mori cone as $M_a = (M_{ai}) \in \cM$, where $a$ labels the generators and $i=1,\ldots, h^{1,1}$ the components of this vector. A vector $K=(K_i)$ then lies in the dual K\"ahler cone $\cK$ if it satisfies the conditions
\begin{equation}\label{eq:dualcone}
M_a \cdot K = \sum_{i=1}^{h^{1,1}} M_{ai} K_i \geq 0\, , 
\end{equation}
for all Mori cone generators $M_a$. Computing the dual of a simplicial cone (the number of generators $M_a$ is equal to the dimension $h^{1,1}$ of the cone) can be performed very quickly, even at large dimensions. The complexity of the problem arises when the Mori cone is non-simplicial, i.e.~the number of generators is larger than $h^{1,1}$. In practice, geometries in the Kreuzer-Skarke database at large $h^{1,1}$ have considerably larger numbers of generators, for instance at $h^{1,1}=100$ there are about $150-200$ generators. One way the exponential scaling of the computational cost now becomes apparent is by scanning over all simplicial subcones of $\cM$: each of these can be dualized efficiently, however, we have to consider roughly $\binom{150}{100}$ such subcones at $h^{1,1}=100$.

\item In this paper we therefore take a more pragmatic approach, and content ourselves with determining a subset of the K\"ahler cone generators. We consider only linearly independent generators, and denote the number of generators we find for our examples in the end by $n_{\rm gen}$. Our method works as follows. We consider a fixed number of simplicial subcones $\cM_{\rm sim}$ of the Mori cone $\cM$, of order $10^{4}$, and dualize each of these individually to a simplicial cone $\cK_{\rm sim}$. Since the subcones we start from are smaller than the Mori cone $\cM_{\rm sim} \subset \cM$, the resulting dual cones are larger than the K\"ahler cone $\cK \subset \cK_{\rm sim}$. In other words, most of the generators of $\cK_{\rm sim}$ do not satisfy all conditions in \eqref{eq:dualcone}. However, typically we do encounter some generators that satisfy \eqref{eq:dualcone}, and we use precisely these rays as generators of the K\"ahler cone. In principle one could recover all generators of the K\"ahler cone $\cK$ with this approach, however around $h^{1,1} \gtrsim 15$ the number of subcones $\cM_{\rm sim}$ is already too large. Nevertheless, by taking only a small subset of Mori subcones we have been able to determine a sizeable number of K\"ahler cone generators for a handful of examples.

\item With the relevant geometric data of the Calabi-Yau hypersurface in hand -- the intersection numbers and the K\"ahler cone generators -- we proceed and study strict asymptotic regimes in the large volume limit. These regions are defined by the ordering of the saxions $s^i$ that specify how far we move along the K\"ahler cone generators. For the procedure to construct the sl(2)-approximation in this strict asymptotic regime we refer to \cite{Grimm:2021ckh} for a pedagogical introduction. One of the main messages we want to convey here is that the resulting boundary Hodge star operator $\star_{\rm sl(2)}$ need not depend on all available moduli -- there can be some trivial sl(2)-triples $(N_i^{\pm}, N_i^0)=0$. By carefully selecting appropriate orderings of the saxions we were able to find sl(2)-approximations depending on about $n_{\rm stab} \sim 20$ moduli in these moduli spaces with $h^{1,1}=70-100$. Note that there are some technical limitations at play here: even for a simplicial K\"ahler cone one would have to consider ($h^{1,1}!$)  different orderings; moreover, we were not able to identify all K\"ahler cone generators but only a subset $n_{\rm gen}< h^{1,1}$. We expect that a complete description of the full non-simplicial K\"ahler cones would enable us to identify asymptotic regimes with $n_{\rm stab}$ much closer to the actual number of moduli $h^{1,1}$. 

\item Given the sl(2)-approximation for a strict asymptotic regime, we proceed and derive lower bounds for the norms $\| G_\ell \|^2_\infty$. We consider real, quantized three-form fluxes $G \in H^3(Y_3, \mathbb{Z})\cap V_{\rm heavy}$ valued in the heavy sl(2)-eigenspaces (see \eqref{eq:Vheavylight}). We subsequently project $G$ onto one of the sl(2)-eigenspaces as $G_{\ell} \in V_\ell \subset V_{\rm heavy}$. Let us note that the sl(2)-splitting is generically only realized over the rationals, so for these individual components of the three-form flux $G$ we have $G_{\ell} \in H^3(Y_3, \mathbb{Q})$. In order to obtain a lower bound on $\|G_{\ell} \|_\infty^2$, we rewrite it as a sum over squares of integer flux quanta: the smallest coefficient in front of these squares then gives a lower bound on the norm. 

\item Let us elaborate on this rewriting of $\|G_{\ell} \|_\infty^2$ for a moment. We take a normalized basis $v_{\ell, a} \in H^3(Y_3, \mathbb{R}) \cap V_\ell$ with $a=1,\ldots , \dim V_\ell$. We represent the $2(h^{2,1}+2)\times \dim(V_\ell)$ matrix of basis vectors by $B_\ell=(v_{\ell,a})$, satisfying $B_\ell^T B_\ell = \mathds{1}_{\dim(V_\ell)}$. Subsequently we can represent $G_\ell$ by a $\dim(V_\ell)$-component vector as $B_\ell^T \cdot G_\ell =(G_{\ell,a})$ such that $G_\ell = G_{\ell,a} v_{\ell,a}$. In this basis the boundary norm can be represented by a $\dim(V_\ell) \times \dim(V_\ell)$ matrix as
\begin{equation}
(M_\ell)_{ab} = \langle v_{\ell, a} , \star_\infty v_{\ell,b} \rangle\, .
\end{equation}
The problem at hand then reduces to decomposing this matrix as $M_\ell = Q^T_\ell Q_\ell$ by a Cholesky decomposition, with $Q_\ell$ a $ \dim(V_\ell) \times \dim(V_\ell) $ matrix: the boundary norm $\|G_\ell \|^2_\infty$ simplifies to the Euclidean norm of a vector as
\begin{equation}
\| G_\ell \|^2_\infty = | Q \cdot B_\ell^T \cdot G_\ell |^2\, .
\end{equation}
Expanding $G$ in an integral basis as $G=(a_I,0)$, where $I=0,\ldots,h^{2,1}$ with $a_I$ integral coefficients, we can then write out this Euclidean norm as
\begin{equation}
\| G_\ell \|^2_\infty = \sum_a  b_a \Bigl( \sum_I n_{aI} a_I\Bigr)^2\, ,
\end{equation}
with $n_{aI} \in \mathbb{Z}$ such that $\text{gcd}_I(n_{aI})=1$ for each $a=1,\ldots, \dim(V_\ell)$, and $b_a \in \mathbb{Q}_{>0}$. The lower bound is then given by
\begin{equation}
\| G_\ell \|^2_\infty \geq \min(b_a)\, .
\end{equation} 

\item Let us demonstrate the above procedure on an example. We take the sl(2)-eigenspace $V_{01_{13}}$ of example 4 in table \ref{table:fluxquant}, which can be spanned by the unit vectors $e_{86}$ and $e_{87}$ with a $1$ in the corresponding positions. The flux $G_{01_{13}}$ along this basis can be expanded as
\begin{equation}
G_{01_{13}} = \frac{1}{3} (-a_{36} - 2 a_{84} + 3 a_{86})e_{86} + \frac{1}{3}(-2a_{36}-a_{84}+3a_{87})e_{87}\, .
\end{equation}
In the basis $e_{86},e_{87}$ the Hodge norm reduces to
\begin{equation}
(M_{0_{12}1_{10}})_{ab} = \begin{pmatrix}
14 & -7 \\
-7 & 14 
\end{pmatrix}, \qquad Q = \begin{pmatrix} \sqrt{14} & -\frac{7}{2} \\
0 & \frac{\sqrt{21}}{2}
\end{pmatrix}\, .
\end{equation}
The norm then simplifies to
\begin{equation}
\|G_{01_{13}}\|^2_\infty = \frac{7}{2} (a_{84}-2a_{86}+a_{87})^2 + \frac{7}{6} (2a_{36}+a_{84}-3a_{87})^2\, ,
\end{equation}
which  has as lower bound
\begin{equation}\label{app_tadpole_001}
\|G_{01_{13}}\|^2_\infty \geq  \frac{7}{6}\, .
\end{equation}
Note, however, that this particular bound is not easily saturated. For instance, by putting only $a_{84}=1$ and all others to zero we also turn on the first square, in which case $\|G_{01_{13}}\|^2_\infty  = 14/3$. This interplay between different squares arises for other sl(2)-eigenspaces as well. It even shows up between different boundary norms $\|G_\ell\|^2_\infty$, where saturating the lower bound for one coefficient results in another satisfying its bound marginally.

\end{itemize}


\bibliographystyle{JHEP}
\bibliography{refs-NET}

\providecommand{\href}[2]{#2}\begingroup\raggedright\begin{thebibliography}{10}

\bibitem{Becker:1996gj}
K.~Becker and M.~Becker, {\it {M theory on eight manifolds}},  {\em Nucl. Phys.
  B} {\bf 477} (1996) 155--167,
  [\href{http://arxiv.org/abs/hep-th/9605053}{{\tt hep-th/9605053}}].

\bibitem{Dasgupta:1999ss}
K.~Dasgupta, G.~Rajesh, and S.~Sethi, {\it {M theory, orientifolds and G -
  flux}},  {\em JHEP} {\bf 08} (1999) 023,
  [\href{http://arxiv.org/abs/hep-th/9908088}{{\tt hep-th/9908088}}].

\bibitem{Grana:2000jj}
M.~{Gra\~na} and J.~Polchinski, {\it {Supersymmetric three form flux
  perturbations on AdS(5)}},  {\em Phys. Rev.} {\bf D63} (2001) 026001,
  [\href{http://arxiv.org/abs/hep-th/0009211}{{\tt hep-th/0009211}}].

\bibitem{Giddings:2001yu}
S.~B. Giddings, S.~Kachru, and J.~Polchinski, {\it {Hierarchies from fluxes in
  string compactifications}},  {\em Phys. Rev.} {\bf D66} (2002) 106006,
  [\href{http://arxiv.org/abs/hep-th/0105097}{{\tt hep-th/0105097}}].

\bibitem{Bena:2020xrh}
I.~Bena, J.~Bl\r{a}b\"ack, M.~Gra\~na, and S.~L\"ust, {\it {The Tadpole
  Problem}},  \href{http://arxiv.org/abs/2010.10519}{{\tt arXiv:2010.10519}}.

\bibitem{Bena:2021wyr}
I.~Bena, J.~Bl\r{a}b\"ack, M.~Gra\~na, and S.~L\"ust, {\it {Algorithmically
  Solving the Tadpole Problem}},  {\em Adv. Appl. Clifford Algebras} {\bf 32}
  (2022), no.~1 7, [\href{http://arxiv.org/abs/2103.03250}{{\tt
  arXiv:2103.03250}}].

\bibitem{Schmid}
W.~Schmid, {\it {Variation of Hodge structure: the singularities of the period
  mapping}},  {\em Invent. Math. , 22:211--319, 1973} (1973).

\bibitem{CKS}
E.~Cattani, A.~Kaplan, and W.~Schmid, {\it {Degeneration of Hodge Structures}},
   {\em Annals of Mathematics} {\bf 123} (1986), no.~3 457--535.

\bibitem{Grimm:2019ixq}
T.~W. Grimm, C.~Li, and I.~Valenzuela, {\it {Asymptotic Flux Compactifications
  and the Swampland}},  {\em JHEP} {\bf 06} (2020) 009,
  [\href{http://arxiv.org/abs/1910.09549}{{\tt arXiv:1910.09549}}]. [Erratum:
  JHEP 01, 007 (2021)].

\bibitem{Grimm:2021ckh}
T.~W. Grimm, E.~Plauschinn, and D.~van~de Heisteeg, {\it {Moduli Stabilization
  in Asymptotic Flux Compactifications}},
  \href{http://arxiv.org/abs/2110.05511}{{\tt arXiv:2110.05511}}.

\bibitem{Denef:2008wq}
F.~Denef, {\it {Les Houches Lectures on Constructing String Vacua}},  {\em Les
  Houches} {\bf 87} (2008) 483--610,
  [\href{http://arxiv.org/abs/0803.1194}{{\tt arXiv:0803.1194}}].

\bibitem{Grimm:2010ks}
T.~W. Grimm, {\it {The N=1 effective action of F-theory compactifications}},
  {\em Nucl. Phys. B} {\bf 845} (2011) 48--92,
  [\href{http://arxiv.org/abs/1008.4133}{{\tt arXiv:1008.4133}}].

\bibitem{Grimm:2013gma}
T.~W. Grimm, R.~Savelli, and M.~Weissenbacher, {\it {On \textbackslash{}alpha'
  corrections in N=1 F-theory compactifications}},  {\em Phys. Lett. B} {\bf
  725} (2013) 431--436, [\href{http://arxiv.org/abs/1303.3317}{{\tt
  arXiv:1303.3317}}].

\bibitem{Grimm:2014efa}
T.~W. Grimm, T.~G. Pugh, and M.~Weissenbacher, {\it {The effective action of
  warped M-theory reductions with higher derivative terms \textemdash{} part
  I}},  {\em JHEP} {\bf 01} (2016) 142,
  [\href{http://arxiv.org/abs/1412.5073}{{\tt arXiv:1412.5073}}].

\bibitem{Grimm:2015mua}
T.~W. Grimm, T.~G. Pugh, and M.~Weissenbacher, {\it {The effective action of
  warped M-theory reductions with higher-derivative terms - Part II}},  {\em
  JHEP} {\bf 12} (2015) 117, [\href{http://arxiv.org/abs/1507.00343}{{\tt
  arXiv:1507.00343}}].

\bibitem{Minasian:2015bxa}
R.~Minasian, T.~G. Pugh, and R.~Savelli, {\it {F-theory at order $\alpha'^3$}},
   {\em JHEP} {\bf 10} (2015) 050, [\href{http://arxiv.org/abs/1506.06756}{{\tt
  arXiv:1506.06756}}].

\bibitem{Weissenbacher:2019mef}
M.~Weissenbacher, {\it {F-theory vacua and $\alpha'$-corrections}},  {\em JHEP}
  {\bf 04} (2020) 032, [\href{http://arxiv.org/abs/1901.04758}{{\tt
  arXiv:1901.04758}}].

\bibitem{Klaewer:2020lfg}
D.~Klaewer, S.-J. Lee, T.~Weigand, and M.~Wiesner, {\it {Quantum corrections in
  4d $N$ = 1 infinite distance limits and the weak gravity conjecture}},  {\em
  JHEP} {\bf 03} (2021) 252, [\href{http://arxiv.org/abs/2011.00024}{{\tt
  arXiv:2011.00024}}].

\bibitem{Cicoli:2021rub}
M.~Cicoli, F.~Quevedo, R.~Savelli, A.~Schachner, and R.~Valandro, {\it
  {Systematics of the \ensuremath{\alpha}' expansion in F-theory}},  {\em JHEP}
  {\bf 08} (2021) 099, [\href{http://arxiv.org/abs/2106.04592}{{\tt
  arXiv:2106.04592}}].

\bibitem{Gao:2022fdi}
X.~Gao, A.~Hebecker, S.~Schreyer, and G.~Venken, {\it {The LVS Parametric
  Tadpole Constraint}},  \href{http://arxiv.org/abs/2202.04087}{{\tt
  arXiv:2202.04087}}.

\bibitem{Haack:2001jz}
M.~Haack and J.~Louis, {\it {M theory compactified on Calabi-Yau fourfolds with
  background flux}},  {\em Phys. Lett. B} {\bf 507} (2001) 296--304,
  [\href{http://arxiv.org/abs/hep-th/0103068}{{\tt hep-th/0103068}}].

\bibitem{Ashok:2003gk}
S.~Ashok and M.~R. Douglas, {\it {Counting flux vacua}},  {\em JHEP} {\bf 01}
  (2004) 060, [\href{http://arxiv.org/abs/hep-th/0307049}{{\tt
  hep-th/0307049}}].

\bibitem{Bakker:2021uqw}
B.~Bakker, T.~W. Grimm, C.~Schnell, and J.~Tsimerman, {\it {Finiteness for
  self-dual classes in integral variations of Hodge structure}},
  \href{http://arxiv.org/abs/2112.06995}{{\tt arXiv:2112.06995}}.

\bibitem{CattaniDeligneKaplan}
E.~Cattani, P.~Deligne, and A.~Kaplan, {\it On the locus of hodge classes},
  {\em Journal of the American Mathematical Society} {\bf 8} (1995), no.~2
  483--506.

\bibitem{Schnell}
C.~Schnell, {\it {Letter to T.~Grimm}},  {\em 2020}.

\bibitem{Grimm:2020cda}
T.~W. Grimm, {\it {Moduli space holography and the finiteness of flux vacua}},
  {\em JHEP} {\bf 10} (2021) 153, [\href{http://arxiv.org/abs/2010.15838}{{\tt
  arXiv:2010.15838}}].

\bibitem{Taylor:2015xtz}
W.~Taylor and Y.-N. Wang, {\it {The F-theory geometry with most flux vacua}},
  {\em JHEP} {\bf 12} (2015) 164, [\href{http://arxiv.org/abs/1511.03209}{{\tt
  arXiv:1511.03209}}].

\bibitem{Bastian:2021eom}
B.~Bastian, T.~W. Grimm, and D.~van~de Heisteeg, {\it {Modeling General
  Asymptotic Calabi-Yau Periods}},  \href{http://arxiv.org/abs/2105.02232}{{\tt
  arXiv:2105.02232}}.

\bibitem{Grimm:2018ohb}
T.~W. Grimm, E.~Palti, and I.~Valenzuela, {\it {Infinite Distances in Field
  Space and Massless Towers of States}},  {\em JHEP} {\bf 08} (2018) 143,
  [\href{http://arxiv.org/abs/1802.08264}{{\tt arXiv:1802.08264}}].

\bibitem{Grimm:2018cpv}
T.~W. Grimm, C.~Li, and E.~Palti, {\it {Infinite Distance Networks in Field
  Space and Charge Orbits}},  {\em JHEP} {\bf 03} (2019) 016,
  [\href{http://arxiv.org/abs/1811.02571}{{\tt arXiv:1811.02571}}].

\bibitem{Grimm:2021ikg}
T.~W. Grimm, J.~Monnee, and D.~van~de Heisteeg, {\it {Bulk Reconstruction in
  Moduli Space Holography}},  \href{http://arxiv.org/abs/2103.12746}{{\tt
  arXiv:2103.12746}}.

\bibitem{Grimm:2021idu}
T.~W. Grimm and J.~Monnee, {\it {Deformed WZW Models and Hodge Theory - Part~ I
  -}},  \href{http://arxiv.org/abs/2112.00031}{{\tt arXiv:2112.00031}}.

\bibitem{Bastian:2020egp}
B.~Bastian, T.~W. Grimm, and D.~van~de Heisteeg, {\it {Weak gravity bounds in
  asymptotic string compactifications}},  {\em JHEP} {\bf 06} (2021) 162,
  [\href{http://arxiv.org/abs/2011.08854}{{\tt arXiv:2011.08854}}].

\bibitem{Bielleman:2015ina}
S.~Bielleman, L.~E. Ibanez, and I.~Valenzuela, {\it {Minkowski 3-forms, Flux
  String Vacua, Axion Stability and Naturalness}},  {\em JHEP} {\bf 12} (2015)
  119, [\href{http://arxiv.org/abs/1507.06793}{{\tt arXiv:1507.06793}}].

\bibitem{Carta:2016ynn}
F.~Carta, F.~Marchesano, W.~Staessens, and G.~Zoccarato, {\it {Open string
  multi-branched and K\"ahler potentials}},  {\em JHEP} {\bf 09} (2016) 062,
  [\href{http://arxiv.org/abs/1606.00508}{{\tt arXiv:1606.00508}}].

\bibitem{Herraez:2018vae}
A.~Herraez, L.~E. Ibanez, F.~Marchesano, and G.~Zoccarato, {\it {The Type IIA
  Flux Potential, 4-forms and Freed-Witten anomalies}},  {\em JHEP} {\bf 09}
  (2018) 018, [\href{http://arxiv.org/abs/1802.05771}{{\tt arXiv:1802.05771}}].

\bibitem{Marchesano:2019hfb}
F.~Marchesano and J.~Quirant, {\it {A Landscape of AdS Flux Vacua}},  {\em
  JHEP} {\bf 12} (2019) 110, [\href{http://arxiv.org/abs/1908.11386}{{\tt
  arXiv:1908.11386}}].

\bibitem{Grimm:2020ouv}
T.~W. Grimm and C.~Li, {\it {Universal axion backreaction in flux
  compactifications}},  {\em JHEP} {\bf 06} (2021) 067,
  [\href{http://arxiv.org/abs/2012.08272}{{\tt arXiv:2012.08272}}].

\bibitem{Betzler:2019kon}
P.~Betzler and E.~Plauschinn, {\it {Type IIB flux vacua and tadpole
  cancellation}},  {\em Fortsch. Phys.} {\bf 67} (2019), no.~11 1900065,
  [\href{http://arxiv.org/abs/1905.08823}{{\tt arXiv:1905.08823}}].

\bibitem{Gendler:2020dfp}
N.~Gendler and I.~Valenzuela, {\it {Merging the weak gravity and distance
  conjectures using BPS extremal black holes}},  {\em JHEP} {\bf 01} (2021)
  176, [\href{http://arxiv.org/abs/2004.10768}{{\tt arXiv:2004.10768}}].

\bibitem{Palti:2008mg}
E.~Palti, G.~Tasinato, and J.~Ward, {\it {WEAKLY-coupled IIA Flux
  Compactifications}},  {\em JHEP} {\bf 06} (2008) 084,
  [\href{http://arxiv.org/abs/0804.1248}{{\tt arXiv:0804.1248}}].

\bibitem{Marchesano:2021gyv}
F.~Marchesano, D.~Prieto, and M.~Wiesner, {\it {F-theory flux vacua at large
  complex structure}},  {\em JHEP} {\bf 08} (2021) 077,
  [\href{http://arxiv.org/abs/2105.09326}{{\tt arXiv:2105.09326}}].

\bibitem{Lust:2021xds}
S.~L\"ust, {\it {Large complex structure flux vacua of IIB and the Tadpole
  Conjecture}},  \href{http://arxiv.org/abs/2109.05033}{{\tt
  arXiv:2109.05033}}.

\bibitem{Plauschinn:2021hkp}
E.~Plauschinn, {\it {The tadpole conjecture at large complex-structure}},  {\em
  JHEP} {\bf 02} (2022) 206, [\href{http://arxiv.org/abs/2109.00029}{{\tt
  arXiv:2109.00029}}].

\bibitem{Sen:1996vd}
A.~Sen, {\it {F theory and orientifolds}},  {\em Nucl. Phys.} {\bf B475} (1996)
  562--578, [\href{http://arxiv.org/abs/hep-th/9605150}{{\tt hep-th/9605150}}].

\bibitem{cytools}
M.~Demirtas, L.~McAllister, and A.~Rios-Tascon, {\it {CYTools: A Software
  Package for Analyzing Calabi-Yau Hypersurfaces in Toric Varieties --- to
  appear}}, .

\bibitem{Kreuzer:2000xy}
M.~Kreuzer and H.~Skarke, {\it {Complete classification of reflexive polyhedra
  in four-dimensions}},  {\em Adv. Theor. Math. Phys.} {\bf 4} (2002)
  1209--1230, [\href{http://arxiv.org/abs/hep-th/0002240}{{\tt
  hep-th/0002240}}].

\bibitem{Altman:2014bfa}
R.~Altman, J.~Gray, Y.-H. He, V.~Jejjala, and B.~D. Nelson, {\it {A Calabi-Yau
  Database: Threefolds Constructed from the Kreuzer-Skarke List}},  {\em JHEP}
  {\bf 02} (2015) 158, [\href{http://arxiv.org/abs/1411.1418}{{\tt
  arXiv:1411.1418}}].

\bibitem{Demirtas:2018akl}
M.~Demirtas, C.~Long, L.~McAllister, and M.~Stillman, {\it {The Kreuzer-Skarke
  Axiverse}},  {\em JHEP} {\bf 04} (2020) 138,
  [\href{http://arxiv.org/abs/1808.01282}{{\tt arXiv:1808.01282}}].

\bibitem{Demirtas:2020dbm}
M.~Demirtas, L.~McAllister, and A.~Rios-Tascon, {\it {Bounding the
  Kreuzer-Skarke Landscape}},  {\em Fortsch. Phys.} {\bf 68} (2020) 2000086,
  [\href{http://arxiv.org/abs/2008.01730}{{\tt arXiv:2008.01730}}].

\end{thebibliography}\endgroup


\end{document}